\documentclass[usenatbib,a4paper,times,fleqn]{mnras}
\usepackage{graphicx,natbib,times,amssymb,amsmath,pstricks}
\usepackage{aas_macros}
\usepackage{bm,url}
\usepackage{textcomp}
\usepackage{breakurl}
\usepackage{subfig}

\newcommand{\newtext}[1]{{\black #1}}

\newcommand {\apgt} {\ {\raise-.5ex\hbox{$\buildrel>\over\sim$}}\ }
\newcommand {\aplt} {\ {\raise-.5ex\hbox{$\buildrel<\over\sim$}}\ }

\title[Rings and gaps in Elias 24]{Rings and gaps in the disc around Elias 24 revealed by ALMA}

\author[Dipierro et al.]{\parbox{\textwidth}{G. Dipierro$^{1,2}$\thanks{giovanni.dipierro@leicester.ac.uk}, L. Ricci$^{3,4,5}$, L. P\'erez$^{6,7}$, G. Lodato$^{2}$, R.~D. Alexander$^{1}$, G. Laibe$^{8}$, \\S. Andrews$^{9}$, J.~M. Carpenter$^{10}$, C. J. Chandler$^{11}$, J.~A. Greaves$^{12}$, C. Hall$^{1}$, T. Henning$^{13}$, W. Kwon$^{14}$,  H. Linz$^{13}$, L. Mundy$^{15}$, A. Sargent$^{16}$, M. Tazzari$^{17}$, L. Testi$^{18,19}$, D. Wilner$^{9}$ }\vspace{0.1cm}\\ 
$^{1}$ Department of Physics and Astronomy, University of Leicester, Leicester, LE1 7RH, UK \\
$^{2}$ Dipartimento di Fisica, Universit\`a degli Studi di Milano, Via Celoria, 16, Milano, I-20133, Italy \\
$^{3}$ Department of Physics and Astronomy, Rice University, 6100 Main Street, MS-108, Houston, Texas 77005, USA \\
$^{4}$ Department of Physics and Astronomy, California State University Northridge, 18111 Nordhoff Street, Northridge CA 91130, USA\\
$^{5}$ Jet Propulsion Laboratory, California Institute of Technology, 4800 Oak Grove Drive, Pasadena, CA, 91109, USA\\
$^{6}$ Universidad de Chile, Departamento de Astronomía, Camino El Observatorio 1515, Las Condes, Santiago, Chile\\
$^{7}$ Max-Planck-Institut f\"ur Radioastronomie, Auf dem H\"ugel 69, D-53121 Bonn, Germany\\
$^{8}$ Univ Lyon, Univ Lyon1, Ens de Lyon, CNRS, Centre de Recherche Astrophysique de Lyon UMR5574, F-69230, Saint-Genis-Laval, France\\
$^{9}$ Harvard-Smithsonian Center for Astrophysics, 60 Garden Street, 02138 Cambridge, MA, USA  \\
$^{10}$ Joint ALMA Observatory, Av. Alonso de Córdova 3107, Vitacura, Santiago, Chile\\
$^{11}$  National Radio Astronomy Observatory, PO Box O, Socorro, NM, USA \\
$^{12}$ Cardiff University, School of Physics \& Astronomy,Cardiff, UK\\
$^{13}$ Max-Planck-Institut f\"ur Astronomie, Heidelberg, Germany\\
$^{14}$ Korea Astronomy and Space Science Institute, Daejeon, Republic of Korea\\
$^{15}$ Department of Astronomy, University of Maryland, College Park, MD, USA\\
$^{16}$ Department of Astronomy, California Institute of Technology, MC 249-17, Pasadena, CA 91125, USA\\
$^{17}$ Institute of Astronomy, University of Cambridge, Madingley Road, CB3 0HA, Cambridge, UK\\
$^{18}$ C2PAP, Excellence Cluster Universe, Ludwig-Maximilians-Universität München, Boltzmannstr. 2, D-85748 Garching, Germany\\
$^{19}$ European Southern Observatory, Karl-Schwarzschild-Str. 2, D-85748 Garching, Germany
}
\pagerange{\pageref{firstpage}--\pageref{lastpage}} \pubyear{2017}
\date{}
\begin{document}
\label{firstpage}
\bibliographystyle{mnras}
\maketitle

\begin{abstract}
We present Atacama Large Millimeter/sub-millimeter Array (ALMA) Cycle 2 observations of the 1.3 mm dust continuum emission of the protoplanetary disc surrounding the T Tauri star Elias 24 with an angular resolution of $\sim 0.2''$ ($\sim 28$ au). 
The dust continuum emission map reveals a dark ring at a radial distance of $0.47''$ ($\sim 65$ au) from the central star, surrounded by a bright ring at $0.58''$ ($\sim 81$ au). In the outer disc, the radial intensity profile shows two inflection points at $0.71''$ and $0.87''$ ($\sim 99$ and $121$ au respectively).
We perform  global three-dimensional smoothed particle hydrodynamic gas/dust simulations of discs hosting a  migrating and accreting planet. Combining the dust density maps of small and large grains with three dimensional radiative transfer calculations, we produce synthetic ALMA observations of a variety of disc models in order to reproduce the gap- and ring-like features observed in Elias 24.
We find that the dust emission across the disc is consistent with the presence of an embedded planet with a mass of $\sim 0.7\, \mathrm{M_{\mathrm{J}}}$ at an orbital radius of $\sim$ 60 au. Our model suggests that the two inflection points in the radial intensity profile are due to the inward radial motion of large dust grains from the outer disc. The surface brightness map of our disc model provides a reasonable match to the gap- and ring-like structures observed in Elias 24, with an average discrepancy of $\sim$ 5\% of the observed fluxes around the gap region.
\end{abstract}

\section{Introduction}
With the advent of the new generation of radio interferometers and improvements in near-infrared imaging, the field of protoplanetary discs is currently being revolutionized. Our ability to image star formation regions has dramatically increased, offering us an unprecedented opportunity to gain insight into protoplanetary disc evolution and planet formation. A recent active topic of research in this field is the characterization of substructures within protoplanetary discs. Among the wide range of morphologies of substructures imaged recently, the most fascinating and elementary ones are gap- and ring-like structures, that might be linked to ongoing and/or a subsequent formation of one or more planets.

Dust rings have been detected in a number of young and evolved protoplanetary discs  by high-fidelity and high-resolution observations both in scattered light emission \citep{de-boer16a,ginski16a,van-boekel16a,pohl17a} and dust thermal emission \citep{alma-partnership15a,canovas16a,andrews16a,isella16b,perez16a,van-der-plas16a,loomis17a,fedele17b,fedele17a,hendler17a}.  
The origin of these structures and their link to the planet formation process is still debated. There are currently two classes of scenarios that have been proposed to explain their origin, differing from one another by the presence of a planet embedded in the disc. 
According to the first scenario, rings originate from self-induced dust pile-ups \citep{gonzalez15a}, dead zones \citep{ruge16a}, rapid pebble growth around condensation fronts \citep{zhang15a}, aggregate sintering \citep{okuzumi15a}, large scale instabilities due to dust settling \citep{loren-aguilar16a}, secular gravitational instabilities \citep{takahashi16a} or large-scale vortices \citep{barge17a}. 
In some of the previous models, the evidence of a gap in the emission disc maps might be the sign of a variation of the optical properties of the dust, which in turn might be related to a variation of the sticking efficiency of dust grains. Other mechanisms in this class of scenario rely on the presence of pressure bumps that can be induced by (magneto)hydrodynamics effects occurring in protoplanetary discs.
In either case, the presence of a gap may indicate grain segregation, which is possibly a sign of a subsequent planet formation \citep{zhang15a}. In this context, by analyzing optically thin continuum observations of HL Tau at 7 mm, \citet{carrasco-gonzalez16a} suggested that the high dust densities measured in the inner rings of HL Tau might be a sign of the formation of gravitationally bound fragments in the rings.

The alternative scenario invokes discs that are dynamically active, in which planets have already formed or are in the act of formation. An embedded planet will excite density waves in the surrounding disc, that then deposit their angular momentum as they are dissipated. If the planet is massive enough, the exchange of angular momentum between the waves created by the planet and the disc results in the formation of a single or multiple gaps, whose morphological features are closely linked to the local disc conditions and the planet properties \citep[e.g.][]{goldreich79a,lin86a,lin93a,crida07a,kley12a,baruteau14a,bae17a,dong17b}.  Recent studies have shown that the resulting gaps carved by Saturn-mass planets can reproduce the structures observed in HL Tau \citep{dipierro15b,jin16a} and HD 163296 \citep{isella16b}. 

From an observational point of view, while the emission observed at a given wavelength most efficiently probes particles of a similar size \citep{draine06a}, the distribution of small dust grains regulates the disc's temperature due to their high absorption opacity at optical and infrared wavelengths. Therefore, the investigation of the dynamics of dust grains with a large range of  sizes in different disc models is crucial to investigate the nature of these structures. 
Importantly, the morphology of the gap in the large and small dust grains is different: small dust grains up to $\sim 10\, \micron$ are expected to be strongly coupled to the gas (except in the very top layers of the disc), so the width and depth of the gap in the micron-sized dust is the same as the gas. By contrast, dust grains in the mm/cm size range experience a significant radial drift and tend to concentrate in regions of pressure maxima that, for massive planets, might be produced at the  gap edges \citep[e.g.][]{lambrechts14a,rosotti16a,dipierro17a}. 
 
In this paper we present Cycle 2 ALMA observations of the protoplanetary disc around Elias 24 in dust thermal emission at a wavelength of 1.3 mm with a resolution of $\sim 0.2'' \times 0.17''$. Elias 24 (also known as WSB 31, YLW 32) is a young ($\sim$ 0.4 Myr), pre-Main Sequence (spectral type of K5) optically visible star \citep{wilking05a}, harbouring one of the brightest millimeter continuum discs in the $\rho$-Ophiucus Star Forming Region (SFR), located at a distance of $139\, \pm\, 6$ pc \citep{mamajek08a}. A better estimate of the distance of the $\rho-$Ophiucus SFR has been recently presented by \citet{ortiz17a}, finding a value of 137.3 $\pm$ 1.2 pc. However, we adopt hereafter the measurement found by \citet{mamajek08a}, that is consistent within one sigma with the value found by \citet{ortiz17a}.
The Spectral Energy Distribution (SED) of the disc around Elias 24 is consistent with a Class II Young Stellar Object  \citep{barsony05a}. Due to its brightness, Elias 24 has been the subject of many observations at sub-millimeter and millimeter wavelengths, especially focused on the inference of disc properties and grain growth processes \citep[e.g.][]{andrews07b,andrews10a,ricci10b}.
In particular, \citet{andrews10a} and \citet{andrews07b} reported observations of Elias 24 with the Submillimeter Array (SMA) interferometer at $880 \,\mathrm{\mu m}$ and 1.3 mm, respectively. These observations revealed a compact disc with a gas mass of the order of $\sim 0.12 \,M_{\odot}$ and a size of  $\sim$ 130 au surrounded by a fainter, extended, and apparently asymmetric emission halo. Both observations did not reveal any substructures at their resolution ($\gtrsim 0.6''$) and were well described by a smooth and axisymmetric distribution of material in the disc  with density monotonically decreasing with distance from the star.
\newtext{While this paper was being refereed, \citet{cox17a} and \citet{cieza17a} published new ALMA continuum observations of Elias 24 at 870~$\mu m$ and 1.3~mm, respectively. The disc was observed adopting shorter integration time (about 45 seconds), producing images with a root mean square noise three times larger compared to the observations shown in this paper.}

The ALMA image presented here reveals a couple of bright and dark rings that might be produced by the combination of dust radial migration and disc-planet interaction. 
We explore this hypothesis by combining global three-dimensional Smoothed Particle Hydrodynamics (SPH) gas and dust simulations of a disc hosting a planet with three-dimensional radiative transfer calculations. The main aim is to reproduce the gap and ring-like structures observed in Elias 24 and provide an estimate of the properties of the planet and the gas and dust disc.

This paper is organized as follows. In Sect.~\ref{sec:almaobs}, we summarize the details of the ALMA observations and imaging. Sect.~\ref{sect:modelling} presents the modelling of the data and the comparison between the expected emission maps from our models and the real ALMA observations. In Sect.~\ref{sect:discussion} we discuss our results and, finally, in Sect.~\ref{sec:conclusion} we summarise our findings and conclusions.

\section{Observations and Data Calibration}
\label{sec:almaobs}

Elias 24  was observed with ALMA during program 2013.1.00498.S (P.I. L. P\'erez), as part of a follow-up of discs with strong long wavelength emission ($\lambda>7$~mm) from the Disks@EVLA collaboration \citep[e.g.][]{perez12a,perez15a,tazzari15a}. ALMA Band 6 observations at a wavelength of 1.3 mm were obtained with 44 antennas, with baselines between 15.1 m to 1.6 km and a total integration time of 12.5 min. The correlator was configured to deliver 6.56 GHz of aggregate continuum bandwidth using eight spectral windows. Five spectral windows were used for continuum observations with coarse spectral resolution, while spectral line observations of $^{12}$CO and isotopologues ($^{13}$CO, C$^{18}$O) were obtained with three additional windows whose line-free channels were used for aggregate continuum bandwidth. The setup is as follows: two wide-bandwidth spectral windows with 1.875 GHz of bandwidth centered at 216.975 and 232.350 GHz, three medium-bandwidth spectral windows with 468.75 MHz of bandwidth centered at 218.777, 219.246 and 231.181 GHz, and three medium-bandwidth spectral windows with 468.75 MHz of bandwidth centered at 230.715, 220.183, and 219.715 GHz, which cover the $J=2-1$ transitions for $^{12}$CO at 230.538 GHz, $^{13}$CO at 220.399 GHz, and C$^{18}$O at 219.560 GHz. 
The observations of $^{12}$CO and $^{13}$CO $J=2-1$ suffered from foreground contamination, while C$^{18}$O $J=2-1$ observations had a low signal-to-noise ratio, these observations will not be further discussed. To calibrate the complex interferometric visibilities the following calibrators were observed: J1517-2422 for bandpass calibration, Titan for absolute surface brightness calibration, and J1627-2426 for gain calibration. Standard calibration was carried out using the ALMA pipeline (\url{https://almascience.nrao.edu/processing/science-pipeline}) inside the Common Astronomy Software Application (CASA) software \citep{mcmullin07a}. 
Imaging of the interferometric observations was carried out with the CASA task \emph{clean} using Briggs weighting of the visibilities with a robust parameter of 0 and multiscale deconvolution \citep{rau11a}, resulting in a root mean square (rms) noise level of $81.7 \,\mathrm{\mu Jy \,beam^{-1}}$. For the multiscale algorithm we selected scales of 0 (point source), 10 and 30 pixels for deconvolution, with a pixel size of $0.02''$. Since the signal-to-noise ratio (SNR) of the continuum observations was high, self calibration (first in phase, then in amplitude) was also carried out, improving the dynamic range of the image from an initial SNR value of $\sim$ 350 to an SNR of $\sim$ 710 in the final image.

\subsection{Continuum emission}
\label{sect:almaobsconti}
\begin{figure}
\begin{center}
\includegraphics[width=0.49\textwidth]{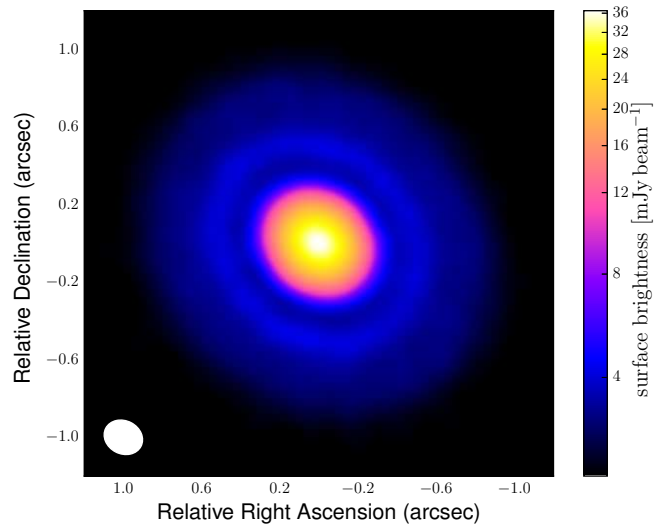}
\caption{ALMA map of the dust thermal continuum emission of the disc around Elias 24 at a wavelength of 1.3 mm. 
The adopted color scale is shown on the right. The white ellipse in the lower left corner indicates the size of the synthesized beam: 0.20'' $\times$ 0.17'', $\mathrm{P.A.} = 62^{\circ}$.} 
\label{fig:continuum}
\end{center}
\end{figure}

Fig.~\ref{fig:continuum}  shows the dust thermal emission map of Elias 24 at the wavelength of 1.3 mm with angular resolution of $0.2'' \times 0.17''$ (P.A. = 62$^{\circ}$), 
corresponding to a physical scale of $28 \times 24$ au at a distance of $139$ pc. 
The dust continuum emission map reveals a bright central core and a dark and bright rings located at $\sim 0.5''$ and $\sim 0.6''$ from the central star, respectively. These features can be considered, to a first approximation \newtext{(the average deviations from axisymmetry are of the order of 0.4\%)}, centrosymmetric around the location of the continuum peak position.
Compared to previous observations of Elias 24 at \mbox{sub-}mm wavelengths, the angular resolution of our observation is three and ten times higher than SMA observations at $880\, \mu m$ and at the same wavelength, respectively \citep{andrews07b,andrews10a}. Moreover, the $880\, \mu m$ SMA continuum observation revealed a bright central core surrounded by extended and asymmetric emission on larger scales \citep{andrews10a}. 
By contrast, Fig.~\ref{fig:continuum} does not show any evidence of a significant deviation from the symmetry around the central peak position in the dust continuum.

We measure the total flux by summing the disc emission with values of the surface brightness within two times the rms noise. 
The measured total flux is $368.1 \pm0.9$ mJy, $\sim$ 10\% larger than the one measured by the previous SMA observation at the same wavelength  \citep{andrews07b,andrews10a}. Comparing the measured total flux with $880\, \mu m$ SMA  emission maps, we obtain a spectral index of $\sim 2.27\pm 0.01$\footnote{The error on the spectral index is computed by propagating the observational uncertainties on the fluxes (e.g. see Eq.~A.5 of \citealt{miotello14a}).}, consistent with the value found by \citet{ricci10b}.
The low value of the spectral index could reflect the presence of regions of high optical depth, which will drive the dust emission spectrum to be close to $F_{\nu}\propto \nu^2$ \citep{ricci12a}, similar to what has been measured in the inner disc region and in the bright rings of HL Tau \citep{alma-partnership15a,carrasco-gonzalez16a,pinte16a}. 
 
In order to infer precise morphological properties of the observed features and the disc on larger scales, we first estimate the disc viewing geometry by analyzing the disc continuum emission map. In detail, we  estimate the disc inclination with respect to our line of sight (defined as $i=0^{\circ}$ for face-on) and its position angle (P.A.), i.e. the amount of disc rotation measured from North to East. We start our analysis by considering an ellipse with geometry parameters found by \citet{andrews10a}, that has been computed by fitting the visibility measurements with those expected by an elliptical Gaussian brightness distribution ($i=24^{\circ}, \mathrm{P.A.}=50^{\circ}$). To refine these parameters, we define a set of elliptical apertures around the location of the continuum peak position and compute the dispersion of the flux measurements inside each of them. We therefore evaluate the best fitting ellipse by computing the elliptical aperture where the spread of the intensity values inside the apertures reaches the minimum value. In other words, we derive the shape of the isophote elliptical contours across all the disc extent.
The weighted average of the best fit inclinations and position angles is $i=28.5^{\circ} \pm 3.8^{\circ}$ and $\mathrm{P.A.}=46.7^{\circ}\pm 6.5^{\circ}$. Both values are consistent with the geometry parameters found by \citet{andrews10a}.

\begin{figure}
\begin{center}
\includegraphics[width=0.48\textwidth]{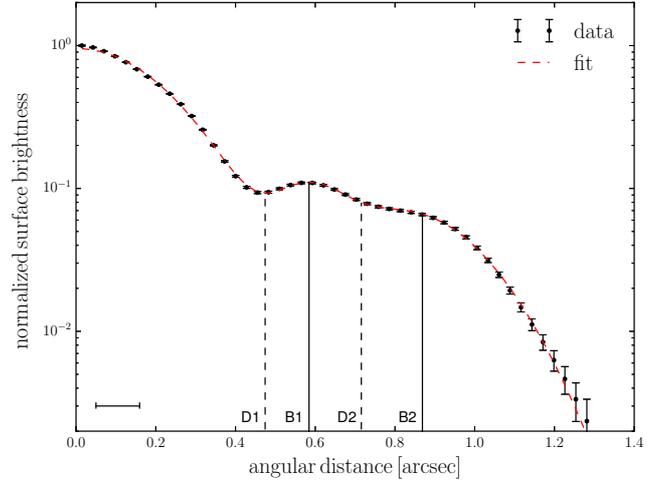}
\caption{Azimuthally averaged profiles of the dust continuum emission (symbols) normalized to the value of the peak intensity. The red dashed line indicates the fitting model based on a combination of Gaussian functions. The error bars are calculated by dividing the rms noise of the observations by the square root of the number of independent beams in each annulus, in order to take into account the increase of the pixel number with the aperture extent. The vertical lines indicate the
position of the dark (D) and bright (B) rings observed in the continuum map, while the horizontal bar in the bottom left corner indicates the angular resolution of the observations, defined as the radius of the circle with the same area as the beam.}
\label{fig:continuum_phot}
\end{center}
\end{figure}

Starting from the measured geometry parameters, we carry out elliptical aperture photometry for a given set of elliptical annuli. Due to the asymmetry of the synthesized beam, we adopt as representative angular resolution for the aperture photometry the radius of the circle with the same area of the beam. The aperture photometry is carried out by considering elliptical annuli with an extent of a quarter of the linear resolution of the observation.
Fig.~\ref{fig:continuum_phot} shows the azimuthally averaged profiles of the dust continuum emission normalized to the peak intensity. The intensity profile shows a pair of dark and bright rings at $\sim \, 0.5''$ and $\sim \, 0.6''$ with a ratio of the continuum intensity of the order of $\sim 0.8$. However, since the angular resolution is not small enough to well resolve the gap along the radial direction (see Fig.~\ref{fig:continuum} and \ref{fig:continuum_phot}), it is reasonably expected that the emission coming from the gap is significantly contaminated by the surrounding  emission.
We note that beyond this pair of rings, the radial intensity profile appears to change concavity at $\sim 0.7''$ (from downward to upward) and $\sim 0.9''$ (from upward to downward) from the central star.
Strictly speaking, the latter two features are very shallow and cannot be classified as real rings, since the radial profile of the intensity appears to monotonically decrease with radial distance. 
However, we include these shallow features in our analysis, since the switch of concavity could indicate a change of dust density distribution induced by the dynamics of large grains.
The two pairs of low and high brightness rings can be fitted by a combination of Gaussian functions to a first approximation. 
We carry out a nonlinear least-squares Levenberg-Marquardt minimization \citep{levenberg44a} taking into account the uncertainties on the flux measurements in order to compute the best fit parameters. Due to the degeneracy on the number of Gaussian functions used for the fit, we fix a combination of three Gaussian functions $f \left(\theta \right)=\sum_{i=1}^3 a_i \exp \left[-\left(\theta-\theta_i \right)^2/2 \sigma_i^2 \right]$, where we assume $\theta_1=0$.
 The values of the best fit parameters are listed in Table~\ref{tab:listfit}, and the comparison between the model and observations is shown in Fig.~\ref{fig:continuum_phot}. The best fit function is the sum of a Gaussian function representing the central bright core with two Gaussian functions centered nearly at the location of the two bright rings.
The fitting function allows better measurement of the radial location of the four features: the dark rings are located at $0.47''$ and $0.71''$ while the bright rings have a semi-major axis of $0.58''$ and $0.87''$, corresponding to physical radii of $65.3$, $98.7$, $80.6$ and $120.9$ au, respectively, assuming a source distance of $139$ pc. The location of the two most extended features is evaluated by finding the change of the concavity of the fitting function of the radial intensity profile. 
The radial profile of the dust continuum emission decreases to the value of the rms noise at $\sim 1.3''$ from the star, which corresponds to a physical outer radius of $\sim 180 $ au.

\newtext{
\citet{cieza17a} reported three gap-like structures in the disc around Elias 24 located at $\sim0.18'', 0.42''$ and $0.7''$, from ALMA observations at 1.3~mm with a spatial resolution of $0.25''\times0.2''$ obtained with a uniform weighting of the visibilities, while \citet{cox17a} reported the presence of one gap in Elias 24 at $\sim 0.42''$ from ALMA observations at $870\,\mu$m with a spatial resolution of $0.21''\times0.18''$ and a natural weighting of the visibilities. 
The singular gap observed by \citet{cox17a} and the second gap observed by \citet{cieza17a} roughly match the gap presented in this paper at $0.47''$, while the inflexion point at $0.71''$ matches with the third gap reported by \citet{cieza17a}. As in \citet{cox17a}, we also do not find evidence for the annular gap at $0.18''$ reported by \citet{cieza17a}.
We note that these previous studies collected observations in ``snapshot'' mode, with about 45s of integration time for each target. This results in sparse uv-coverage and lower image fidelity compared to the observations presented here, which have $\times17$ longer integration time, a factor of 3 better sensitivity, and better spatial resolution, and thus provide a more accurate determination of the structures of the disc around Elias 24.}

\section{Modelling the continuum dust emission}
\label{sect:modelling}

We model the dust continuum emission by assuming that the observed gap corresponds to a dip in the dust density distribution of large grains induced by the presence of an embedded perturber.  
We explore this hypothesis by performing  3D simulations of a variety of dust/gas disc models hosting an embedded protoplanet. The resulting dust density distribution is taken as a disc model for radiative transfer simulations in order to compute the model synthetic images for the dust thermal emission at the wavelength of the observations. Finally, in order to compare the numerical results with the ALMA data, we simulate  observations of our disc models with similar characteristics as our ALMA observations.

Our aim is to obtain an estimate of the main physical properties of a planet able to carve such a gap in the dust density distribution of millimetre grains and infer the disc conditions that might produce the shallow ring-like feature in the outer disc. Given the large number of parameters involved in our analysis and the computationally expensive approach, the fitting procedure is done by eye without performing any statistical test to quantify the goodness of the fit.  
\begin{table}
\begin{center}
\begin{tabular}{|c|c|}
Parameter & Value\\
\hline
$a_1$ & $0.955{\color{white}}\pm0.005$ \\
$\sigma_1$ & $0.194''\pm0.001''$ \\
\hline
$a_2$ & $0.072{\color{white}}\pm0.007$ \\
$\theta_2$ & $0.578''\pm0.004''$ \\
$\sigma_2$ & $0.086''\pm0.006''$ \\
\hline
$a_3$ & $0.069{\color{white}}\pm0.002$ \\
$\theta_3$ & $0.816''\pm0.017''$ \\
$\sigma_3$ & $0.173''\pm0.011''$ \\
\hline
\end{tabular}
\caption{Best fit values of the parametric model of the radial profile of the dust continuum emission computed by carrying out a nonlinear least-squares Marquardt-Levenberg minimization accounting for the uncertainties on the flux measurements.}
\label{tab:listfit}
\end{center}
\end{table}%

\subsection{Methods}
\subsubsection{Dust/gas numerical simulations}
We  perform a set of 3D SPH simulations of gas and dust disc with an embedded protoplanet using the \textsc{phantom} code \citep{price17a}. We approximate the dust disc as a two component system in terms of dust species: a population of micron-sized grains perfectly mixed with the gas and responsible for the thermal structure of the disc and a population of millimetre-sized dust grains whose spatial distribution regulates the disc appearance at millimetre wavelengths. Since our ALMA Band 6 image essentially traces particles with a size comparable to the observing wavelength \citep{draine06a}, we focus our analysis on the dynamics of millimetre dust grains. Due to the tight aerodynamical coupling between micron sized grains and the gas, we do not perform any simulation with smaller dust grains and assume that their 3D spatial density distribution matches that of the gas.

In the \textsc{phantom} code the dust dynamics can be computed using two different approaches:  the two fluid algorithm described by \citet{laibe12b,laibe12a}, typically used for large grains in weak drag regime (i.e. large Stokes numbers) and the one-fluid algorithm \citep{laibe14a,price15a} based on the terminal velocity approximation \citep[e.g.][]{youdin05a} and best suited to simulate particles tightly coupled with the gas (i.e. small Stokes numbers). In terms of dust-gas modelling, the two fluid algorithm treats the dust and gas as two interacting  fluids while in the one-fluid approach the SPH particles represent the whole gas-dust mixture and the dust fraction is evolved as a local property of the mixture.
Both algorithms have been extensively benchmarked on simple test problems including waves, shocks and planets in dust/gas mixtures \citep{laibe12a,price15a,dipierro17a,price17a}. 
In most of our simulations millimetre dust particles are characterized by a Stokes number of the order of $\sim \,0.01$ \newtext{in the midplane, increasing to values close to $\sim$ 0.08 in the disc surface layers at $2H_{\mathrm{g}}$, where $H_{\mathrm{g}}$ is the gas pressure scale height}. Therefore, we adopt the one-fluid algorithm to compute the dust evolution. Moreover, as described by \citet{price15a}, we adopt an alternative implementation of the one fluid algorithm based on the evolution of the quantity $\sqrt{\rho \epsilon}$, where $\rho = \rho_{\rm g} + \rho_{\rm d}$ is the total density of the mixture and $\epsilon$ is the dust fraction $\epsilon \equiv \rho_{\rm d}/\rho$. This new implementation gives better results in  the outer disc regions and the upper surface layers where the dust radial inward motion and settling reduce the dust fraction \citep{price15a,hutchison16a}.

In all our simulations we set the dust grain size to be $s_{\rm grain}=1\, {\rm mm}$ in order to focus our analysis on the dust density distribution probed by the ALMA observations. 
 For the adopted disc models, millimetre grain sizes fall into the Epstein regime, implying that the Stokes number is given by \citet{price15a}
\begin{equation}
\mathrm{St} = \sqrt{\frac{\pi}{8}}\frac{ \rho_{\rm grain} s_{\rm grain} \Omega_{\mathrm{k}}}{\rho c_{\rm s}f}  ,
\label{eq:ts}
\end{equation}
where $\rho_{\rm grain}$ is the intrinsic grain density, $c_{\rm s}$ is the sound speed, $\Omega_{\mathrm{k}}$ is the Keplerian angular velocity and $f$ is a correction factor for supersonic drag. 
We represent the protoplanet and the central star using sink particles \citep{bate95a}. The sinks are free to migrate due to their interaction with the disc and are able to accrete gas and dust particles \citep{bate95a}.


 \subsubsection{Radiative transfer and synthetic ALMA observations}
 \label{sec:rf}

We computed synthetic ALMA observations of our models using the RADMC-3D Monte Carlo  radiative transfer code \citep{dullemond12a} together with the CASA ALMA simulator (version 4.7.1), focusing on ALMA Band 6 (continuum emission at 230 GHz). The main inputs for the radiative transfer modelling are the dust density structure of large and small grains, a model for the dust opacities and the source of luminosity. 
We adopt the dust model used by \citet{andrews10a}, assuming an InterStellar Medium (ISM) composition \citep{draine84a,weingartner01a} with a power-law grain size distribution given by $n\left(s_{\rm grain}\right) \propto s_{\rm grain}^{-m}$ between $0.005\, \mathrm{\mu m}$ and $1$ mm, with $m=3.5$.

The first step is to run radiative transfer simulations to compute the dust temperature as the result of a balance between radiative absorption and re-emission assuming that the star is the only source of luminosity. We model the emission of the central source as a black body, using the star properties adopted by \citet{andrews10a}, i.e. a pre-main sequence star of spectral type K5, radius of $4.2 \,R_{\odot}$ and an effective fixed temperature of 4250 K \citep{wilking05a}. Our analysis does not include the gas as a source of radiation and assumes that the gas temperature only influences the disc vertical geometry.
 
We then create full-resolution images by performing ray-tracing using $10^8$ photon packages. For the comparison with the results of the ALMA observations of Elias 24 we first applied to our models the same disc inclination and position angle as constrained for the Elias 24 disc (Sect.~\ref{sec:almaobs}).
We then calculated the predicted visibilities from our models by Fourier transforming the model images and sampling the visibility function on the same $\left(u,v\right)$ points as probed by our ALMA observations. We then imaged the model visibilities using the same procedure outlined in Section~\ref{sec:almaobs} for the imaging of the ALMA observed visibilities. In this way, the resulting model map has the same characteristics, in terms of angular resolution and sensitivity to emission on different angular scales, as the observed map. 
For each model, the disc was centered to the position in the sky as measured at the brightest pixel in the observed ALMA map. This was done using the \texttt{fixvis} task in CASA.

\subsection{Initial conditions}
\label{sect:initcond}
Due to the large number of degrees of freedom in our modelling, we fix a subsample of the initial disc parameters: the properties of the central star \citep{wilking05a}, the total dust mass and the vertical geometry of the gas disc. The free parameters in the initial disc properties are therefore the axisymmetric surface density profile of both phases and the perturber properties.
Our reference model for the total dust mass (see Sect.~\ref{sect:refmodel}) and the disc vertical geometry are taken from \citet{andrews10a}, following the formalism introduced by \citet{andrews09a}. 

The system consists of a central star of mass $1\, M_{\odot}$ \citep{wilking05a} surrounded by a gas and dust disc extending from $r_{\mathrm{in}}$ = 1 au to $r_{\mathrm{out}}$ = 180 au and modelled as a set of $10^{6}$ SPH gas/dust particles. 
The disc is vertically extended by assuming Gaussian profiles for the volume density in the vertical direction with a thickness of $H_{\mathrm{g}}\left(r\right)=H_{\rm in} \left (r/r_{\rm in}\right)^{1.03}$ where $H_{\rm in}=0.075$ au is the height at the inner radius, $r_{\mathrm{in}}$. The value of the flaring index has been inferred by \citet{andrews10a} by fitting the observed continuum visibilities and the broadband SED of the disc around Elias 24.
We adopt a vertically isothermal equation of state $P=c_{\rm s}^{2} \rho_{\mathrm{g}}$ with $c_{\rm s}\left(r\right) = c_{{\rm s,in}} (r/r_{\rm in})^{-q}$ with a value $ c_{{\rm s,in}}$ and $q$ computed by assuming vertically hydrostatic equilibrium across the disc, i.e. $H_{\mathrm{g}}=c_{\rm s}/\Omega_{\mathrm{k}}$.  
 We set the SPH $\alpha_{\rm AV}$ viscosity parameter to 0.1, resulting in an effective \citet{shakura73a} viscosity $\alpha_{\mathrm{SS}} \approx 0.003$ \citealt{stoll14a,meheut15a}.
 
We run simulations adopting different planet mass and radial distance from the central star. 
The ring-like features at $\sim 0.6''$ from the central star (see Sect.~\ref{sec:almaobs}) might suggest that this structure is a trap of millimetre sized grains due to the presence of a pressure maximum at the outer edge of the gap carved by the planet.
Therefore, we consider a range of planet mass around the expected value able to affect the gas pressure structure, i.e. $M_{\mathrm{p}} \sim \,0.19\, \mathrm{M_{\mathrm{J}}}$ (see Eq.~\ref{eq:mlimgas} below).
 Since we expect a radial inward migration of the planet due to its interaction with the disc, we run simulations by locating the planet far from the central star with respect to the position of the minimum of the intensity profile in the gap region (i.e. $r_{\mathrm{p}} \gtrsim 65$ au).
 Each simulation is evolved for a maximum time of 100 orbits at the initial planet distance from the central star (corresponding to $\sim \,5.2 \times 10^4$ years). However, we analyze the result of the simulations at the end of every orbit in order to find the best model  that matches the observations.

\subsubsection{Models for the disc surface density structure}
\label{sect:refmodel}
We explore different initial disc conditions by varying the shape of the gas and dust surface density profiles, adopting the same dust mass and the disc vertical geometry found by \citet{andrews10a}.
In the model presented by \citet{andrews10a}, the dust and gas surface density profile are assumed to be power laws with an exponential taper at large radii, i.e.
\begin{equation}
\Sigma_{\mathrm{d}}(r)=\Sigma_{\mathrm{c}} \left(\frac{r}{r_{c}} \right)^{-p} \exp \left[- \left(\frac{r}{r_{c}} \right)^{2-p}  \right ],
\label{eq:sigmad}
\end{equation}
where the normalization factor $\Sigma_{c}$ can be computed by integrating Eq.~\ref{eq:sigmad} over the disc (for $p \neq 2$), 
\begin{equation}
\Sigma_{\mathrm{c}}=\left( 2-p \right) \frac{M_{\mathrm{d}}}{2\pi r_{\mathrm{c}}^2}  \exp \left[-\left(\frac{r_{\mathrm{in}}}{r_{c}} \right)^{2-p}  \right ],
\label{eq:sigmad}
\end{equation}
where $M_{\mathrm{d}}=0.0017 M_{\odot}$ is the dust disc mass, $r_{\mathrm{c}}=127$ au is a characteristic scaling radius and $p=0.9$ is the gradient parameter.  
Gas and dust are assumed to be initially well mixed across the disc, with a total gas mass scaled by a constant factor given by the typical ISM dust-to-gas ratio, $\epsilon=0.01$ \citep{mathis77a}. 
The motivation behind this approach is that the gas and dust surface density profile are assumed to be described by a class of solutions of the gas viscous diffusion equation assuming a static (constant with
time) viscosity that follows a simple power-law dependence with radius, $\nu\propto r^{p}$ \citep{lynden-bell74a} and requiring that the system is older than the viscous timescale \citep{hartmann98a,lodato17a}.  The latter assumption seems not to be the case for Elias 24, due to the relatively young age of the system ($\sim$ 0.4 Myr).
However, while the adoption of this gas surface density is physically motivated,  the assumed profile for the dust density structure is physically inaccurate due to the strongly different dynamics between large dust grains and gas. 
To simply account for this kind of deviation from the standard approach, in our modelling we also include pure power law surface density profiles, i.e. without the tapering at  $r_{\mathrm{c}}$ shown in equation \ref{eq:sigmad}, and different dust-to-gas ratios across the disc.
However, as we show below, the tapering in the outer disc region is a natural consequence of the dynamics of large grains. Our simulations produce at the late stages nearly the same shape of the dust density distribution in the outer disc region, regardless our choice for the shape of the initial surface density in the outer disc (between power law or tapered power law). 

 \subsubsection{Dust-to-gas mass ratio}
One of the key parameters in disc modelling is the mass of the gas. The local gas surface density regulates the aerodynamical coupling between the gas and the dust and the accretion and migration efficiency of the planet. The typical approach is to assume a fixed value across all the disc given by the typical ISM dust-to-gas ratio of 0.01  \citep{mathis77a}. 
However, recent high sensitivity and high resolutions disc observations have revealed a discrepancy between dust and gas disc sizes, showing also that the maximum size of dust grains is a function of distance from the star \citep{panic09a,andrews12a,perez12a,rosenfeld13a,de-gregorio-monsalvo13a,perez15a,guidi16a,canovas16a,tazzari15a,carrasco-gonzalez16a}.
 
 This discrepancy might be explained by a combination of radial drift, grain growth processing and differences in optical depth of the two phases \citep[e.g.][]{dutrey98a,birnstiel14a,facchini17a}. In most  invoked scenarios, the location of the outer edge in the dust radial distribution closely depends on the local gas and dust properties, particularly the gas volume density, the dust grain size, the local pressure profile, the turbulence and the dust-to-gas ratio. 
In most cases, using radial drift or optical depth effects to account for the different appearance of the disc in different tracers, the gas outer radius probed by $^{12}$CO emission might differ by a factor of $2-4$ from the dust outer radius probed by millimetre emission \citep{birnstiel14a,facchini17a}.
It is therefore reasonable to expect that the local (i.e. in the dust disc) dust-to-gas ratio is larger than the typical 1\% of the interstellar medium, as recently found in observations of $\mathrm{CO}$ isotopologues emission line and dust continuum \citep[e.g.][]{panic08a,pinte16a,ansdell16a}. 
However, the conversion from CO molecular emission to $\mathrm{H_2}$ mass is hampered by a large range of uncertainties in the exact values of local density and temperature in gas and dust phase, which strongly affect the chemical reactions \citep{qi15a}, leading to a significant overestimation of the dust-to-gas ratio \citep{miotello17a}.

Using simple physical arguments, a rough estimate of the dust-to-gas ratio enhancement induced by radial drift can be obtained by comparing the radial extent of the gas and dust disc. Let us assume that  dust-to-gas ratio in millimetre grains of the disc at birth (when the gas and dust were well-mixed) is $\epsilon_{0}=10^{-2}$. 
We assume that at the age of the disc the outer radius of the gas $r_{\mathrm{g,out}}$ is $\sim2-4$ times larger than the outer radius of the millimetre dust disc $r_{\mathrm{d,out}}$ due to the radial dust drift. We caution that this approach is too simplified due to the large number of caveats involved but allows us to have an estimate of the resulting dust-to-gas mass ratio.
The resulting dust-to-gas ratio in the dusty disc due to the radial dust drift is given by
\begin{equation}
\epsilon_{\mathrm{drift}}=\epsilon_{0}\left (\frac{r_{\mathrm{g,out}}}{r_{\mathrm{d,out}}}\right )^2  \approx 0.04-0.16.
\end{equation}
Moreover, long-wavelength observations of Elias 24 have shown that the spectral index of the millimeter dust opacity is lower than the typical value found for the ISM \citep{ricci10b}. This is strong evidence for the presence of large ($\sim$ mm) dust grains in the outer disc regions. Importantly, the SED fitting shown by \citet{ricci10b}  is consistent with a power-law index of the grain size number density $m=3$, suggesting that the dust mass in the Elias 24 disc is dominated by the largest grains in the dust population, likely with sizes of the order of $\sim$ 1 mm.
\begin{figure*}
\begin{center}
\includegraphics[width=0.4\textwidth]{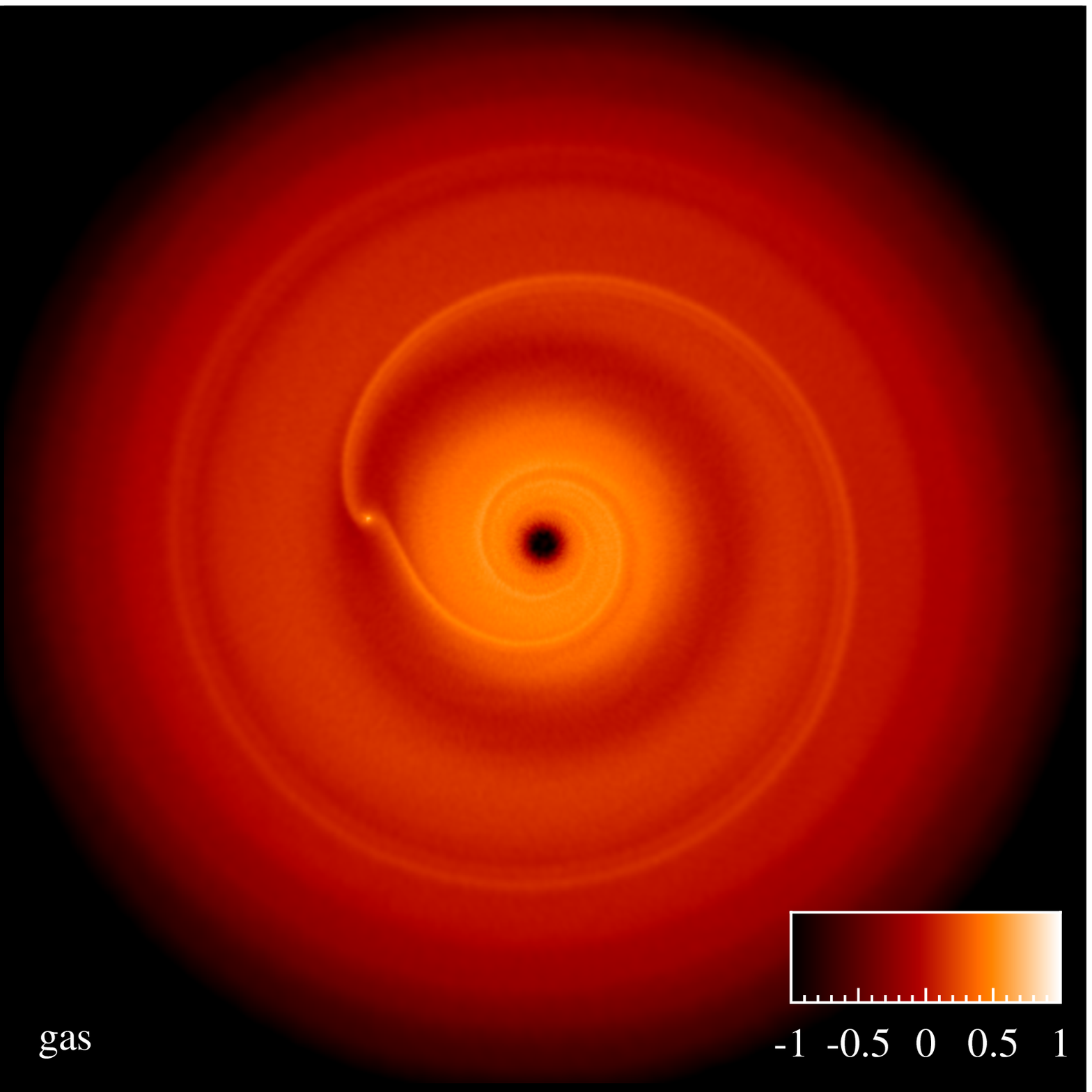}
\includegraphics[width=0.4\textwidth]{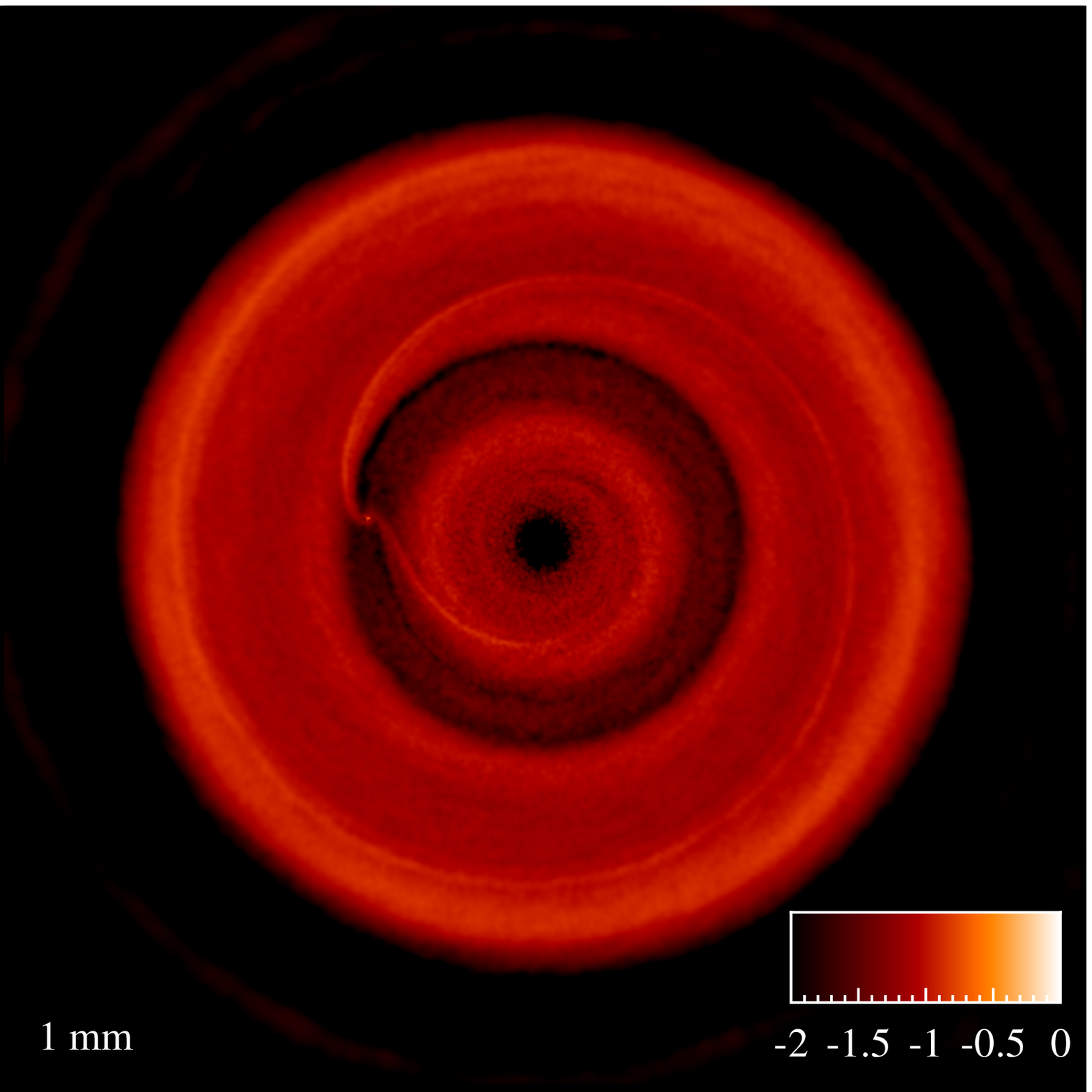}
\caption{Rendered images of gas (left) and millimetre dust grain (right) surface density (in units of $\mathrm{g\,cm^{-2}}$ on a logarithmic scale) of a dusty disc  hosting a planet with an initial mass of $0.15 \, \mathrm{M_{\mathrm{J}}}$ and initially located at a distance of 65 au from the central star.}
\label{fig:renderingsigmagd}
\end{center}
\end{figure*}
 
An additional reason why we adopt a different value for the initial dust-to-gas ratio is that the gas model for Elias 24 presented by \citet{andrews10a} shows a relatively high disc-to-star mass ratio of $\sim 0.12$, which might indicate a gravitationally unstable gas disc. Adopting the same disc model as by \citet{andrews10a}, we compute the stability parameter with respect to gravitationally instabilities \citep{toomre64a}
\begin{equation}
Q\equiv\frac{c_{\mathrm{s}} \kappa}{\pi \mathcal{G} \Sigma_{\mathrm{g}}},
\label{eq:q_def}
\end{equation}
where $\mathcal{G}$ is the gravitational constant \newtext{and $\kappa$ is the epicyclic frequency (equal to $\Omega_{\mathrm{k}}$ in Keplerian discs)}.
We found that in the outer disc, the value of the $Q$-parameter reaches values close to 2, which is close to the expected limiting $Q$-value for the onset of gravitational instability in geometrically thick discs \citep{kratter16a}. In this case, the disc would show the development of large-scale density fluctuations in the form of a spiral pattern with a low number of spiral arms \citep{lodato05a,cossins09a} which should be unambiguously identified through \mbox{(sub-)}millimetre observations \citep{dipierro14a,dipierro15a,dong15c,hall16a}. 
Therefore, the absence of any deviation from axisymmetry might suggest that the gas disc might be not as massive as shown by \citet{andrews10a}.
In our modelling, we therefore adopt a total initial dust-to-gas ratio for millimetre grains of 0.1. 
 
\subsection{Results}

We have carried out a large number of disc+planet gas/dust simulations in order to determine the disc models that provide a satisfactory match with the radial intensity profile of the dust continuum emission as observed by ALMA. 
We caution that the gas and dust density distributions at the end of our simulations are not in a steady state. This is related to the fact that in our simulations the planet is allowed to migrate and accrete due to the tidal interaction with the surrounding disc. As we will describe below, the planet migration and accretion lead to the formation of an unsteady gas and dust density structure around its orbit. If we assume that the shallow gap observed by ALMA corresponds to  a shallow gap in the dust density distribution created by a low mass planet, we can reasonably expect that the planet is still largely embedded in the protoplanetary disc, migrating and accreting mass due to the large gas density around its location. Therefore, since the planet mass and location are not fixed in our simulations, the gas and dust density will evolve accordingly. 
Due to the fact that the gas and dust density are not in steady state, we have chosen a simulation snapshot that best fits the data.
It is therefore reasonable to expect a degree of degeneracy with the simulation parameters. Given the large range of parameters involved in the fitting procedure, we cannot reasonably quantify this degeneracy (see Sect.~\ref{sect:models}). 

In any case, although our approach introduces degeneracies in the disc model parameters, our simulations show a good match to the observation in a small range of disc model parameters around the ones that best reproduce the data, especially regarding the gas mass and the shape of the gas surface density profile.  
In this section we present the disc structure and the simulated ALMA observation of our best model.
\subsubsection{Disc structure}
\label{sect:resulstruct}
%
Our best model consists of an initial gas and dust surface density with a power law function with radius, $\Sigma \propto r^{-0.7}$, hosting a planet with an initial mass of $0.15 \, \mathrm{M_{\mathrm{J}}}$, initially located at a distance of 65 au from the central star. 
We stop the simulations after $\sim$ 85 orbits at the initial location of the planet (corresponding to $\sim \,4.4 \times 10^4$ years) and we use the resulting dust surface density for radiative transfer simulations. At this stage of the simulation, the planet is at 61.7 au from the central star and its mass is $0.7\, \mathrm{M_{\mathrm{J}}}$. 
The surface density structure of both phases (see Fig.~\ref{fig:renderingsigmagd}) shows an annular gap around the planet location and a spiral structure across the disc. Due to the tight coupling between the gas and dust phase in most of the disc, these features are common in both phases. 

\begin{figure}
\begin{center}
\includegraphics[width=0.52\textwidth]{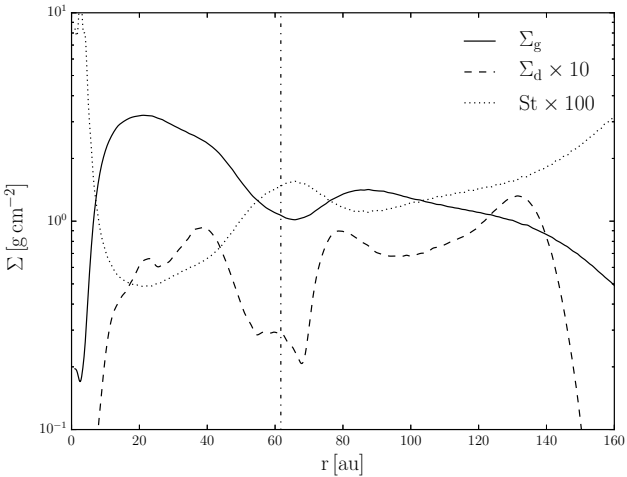}
\caption{Azimuthally averaged dust (dashed) and gas (solid) surface density for our best disc hosting a planet with an initial mass of $0.15 \,\mathrm{M_{\mathrm{J}}}$, initially located at 65 au from the central star. The dust surface density is scaled by a factor of 10 at the end of the simulation, for direct comparison with the gas phase. After 85 orbits (measured at 65 au), the planet is at 61.7 au from the central star and its mass is $0.7\, \mathrm{M_{\mathrm{J}}}$. 
The dotted line indicates the Stokes number multipled by 100. The vertical line indicates the planet orbit.}
\label{fig:sigmagd}
\end{center}
\end{figure}

Fig.~\ref{fig:sigmagd} shows the azimuthally averaged surface density of the gas and dust at the end of the simulation. It can be noticed that the gap in the gas density structure is very shallow: the planet's co-orbital surface density drops to only $\sim$ 60\% of its initial value.  
As expected,  the gap depth in the dust is larger than the one in the gas due to both the higher efficiency of the tidal torque and the dust radial motion induced by drag \citep{paardekooper04a,paardekooper06a,fouchet07a,dipierro16a}. In detail, our simulation shows that the planet is able to perturb the local pressure profile and create a pressure maximum at the gap outer edge (see Sect.~\ref{sect:gapopening}). As a result, mm grains accumulates at the location of the pressure maximum, forming a deeper dust gap than in the gas. At the inner edge, we find that the perturbation of the pressure profile induced by the tidal torque does not exceed the background pressure gradient, leading to a weak pile up at the inner gap edge and an accelerated radial inflow toward the central star \citep{crida07a,fouchet10a}. 

\begin{figure*}
\begin{center}
\includegraphics[height=0.307\textwidth]{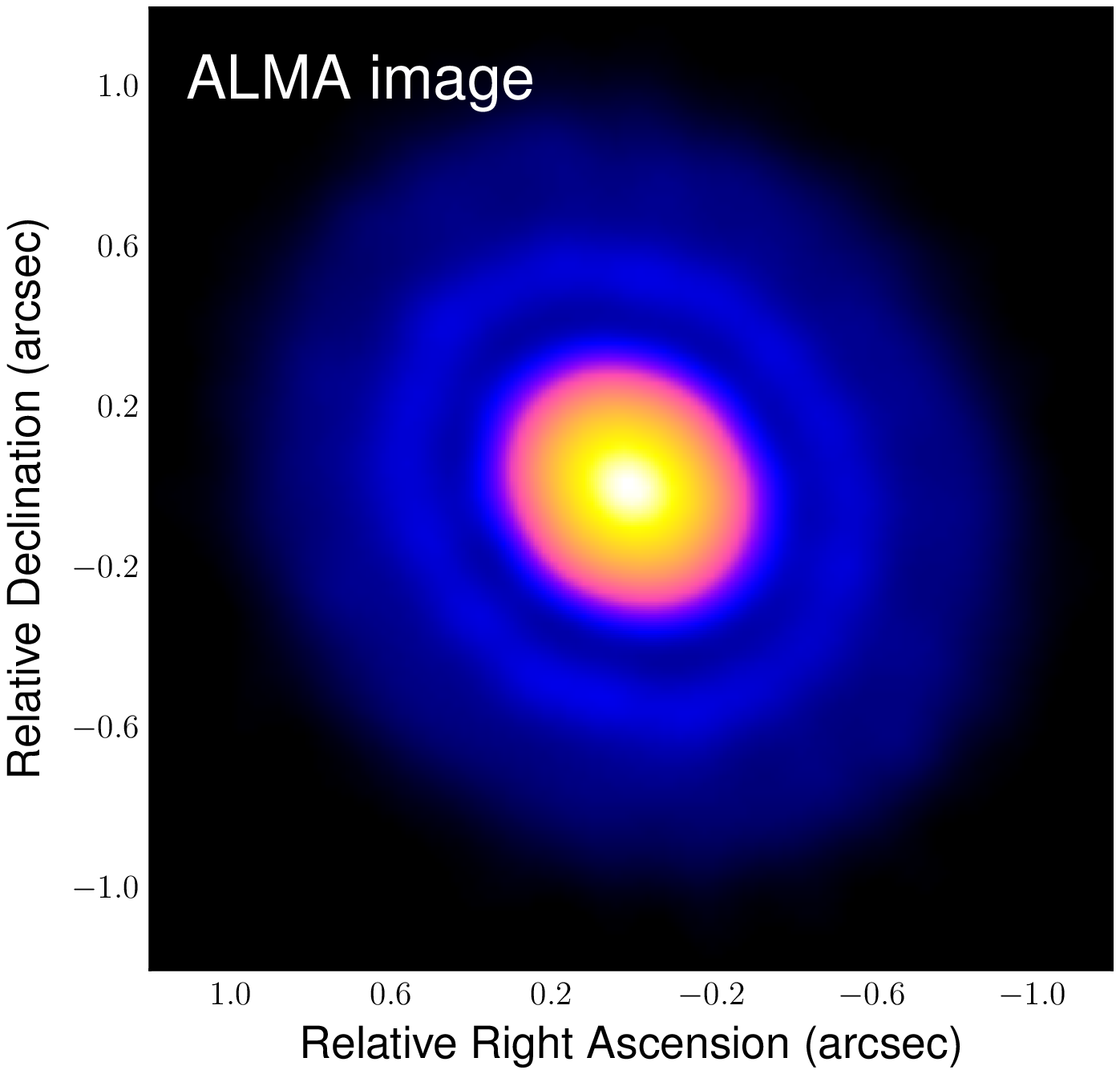}
\includegraphics[height=0.309\textwidth,trim={1.65cm 0 0.36cm 0},clip]{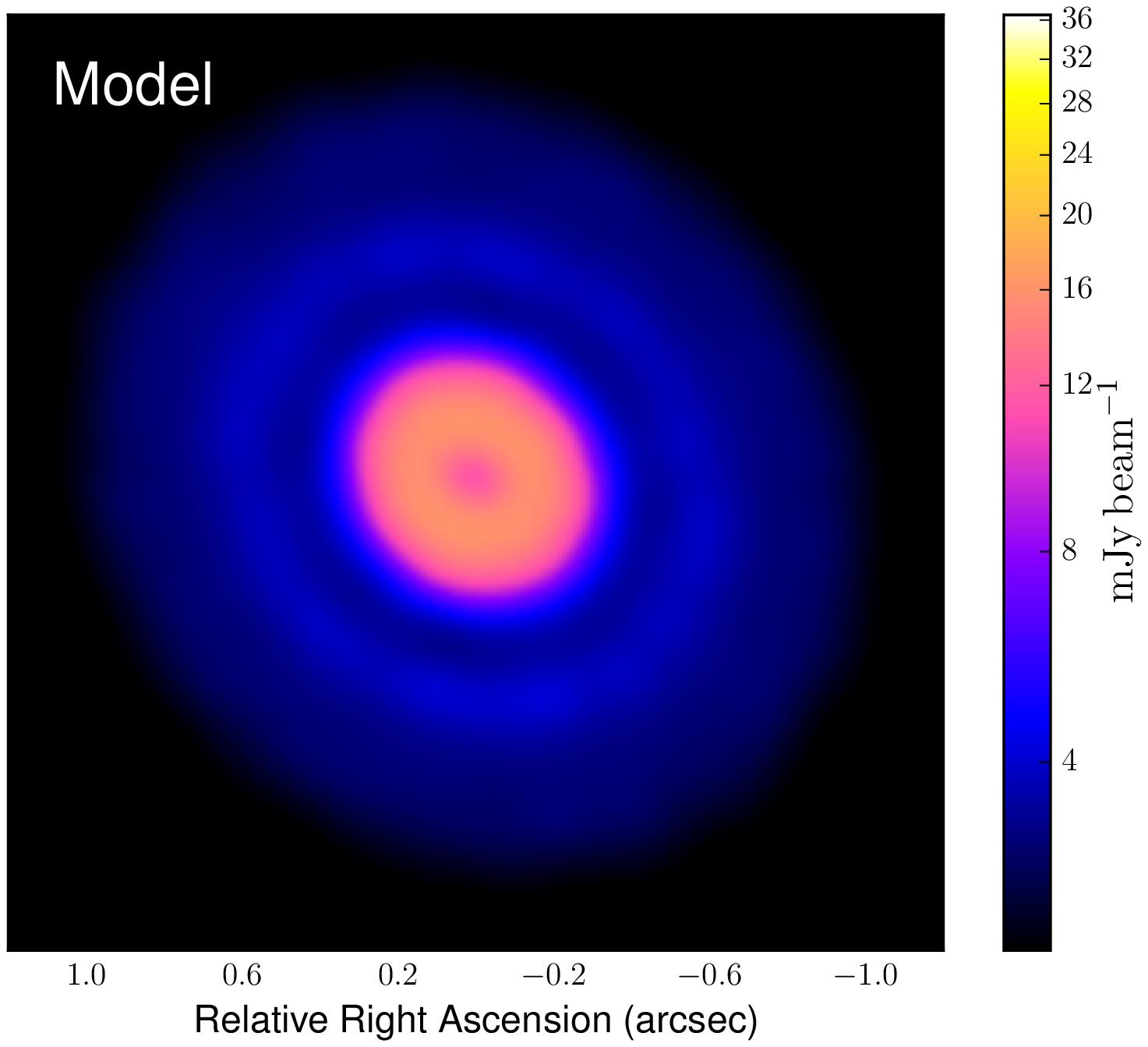}
\includegraphics[height=0.307\textwidth,trim={1.65cm 0 0.36cm 0},clip]{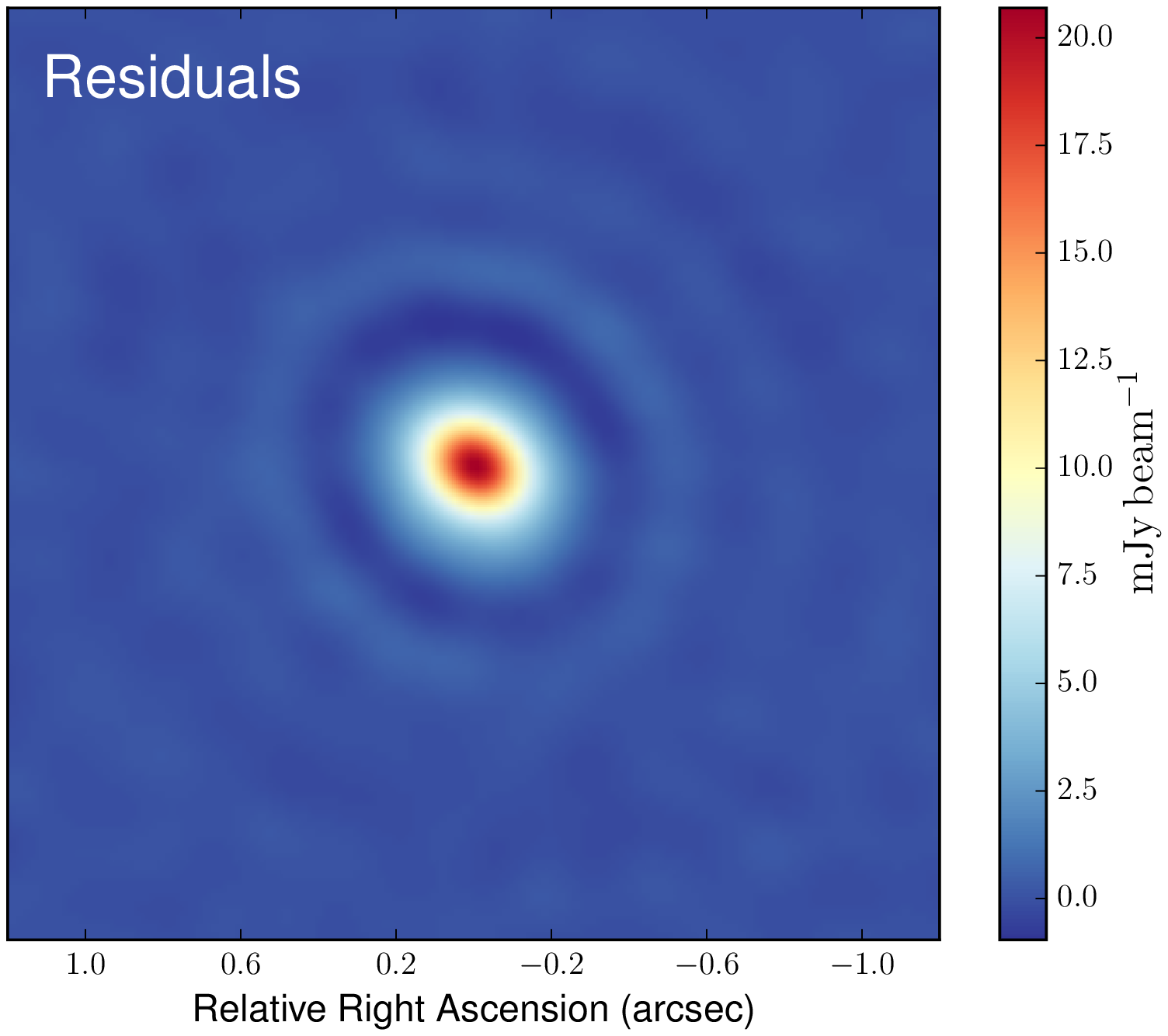}
\caption{Comparison between the ALMA map of Elias 24 (left) and the simulated observation of our best model (middle). The right panel shows the map of the data $-$ model residuals, obtained by imaging the residuals for the real and imaginary parts of the visibility function on the same $\left(u,v\right)$ points as in the ALMA observations.} 
\label{fig:cfrmodel}
\end{center}
\end{figure*}
\begin{figure}
\begin{center}
\includegraphics[width=0.48\textwidth]{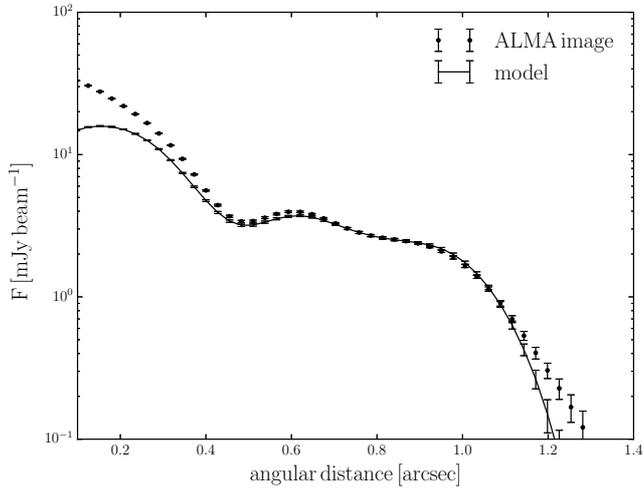}
\caption{Comparison of the surface brightness of our model and the real ALMA observation
using a line cut along the major axis of the disc. In an annulus between 0.4'' and 1'' from the central star, the mean value of the absolute values of the residuals is $\sim 5$\% of the observed fluxes.}
\label{fig:cfrphotom}
\end{center}
\end{figure}
\begin{figure}
\begin{center}
\includegraphics[width=0.465\textwidth]{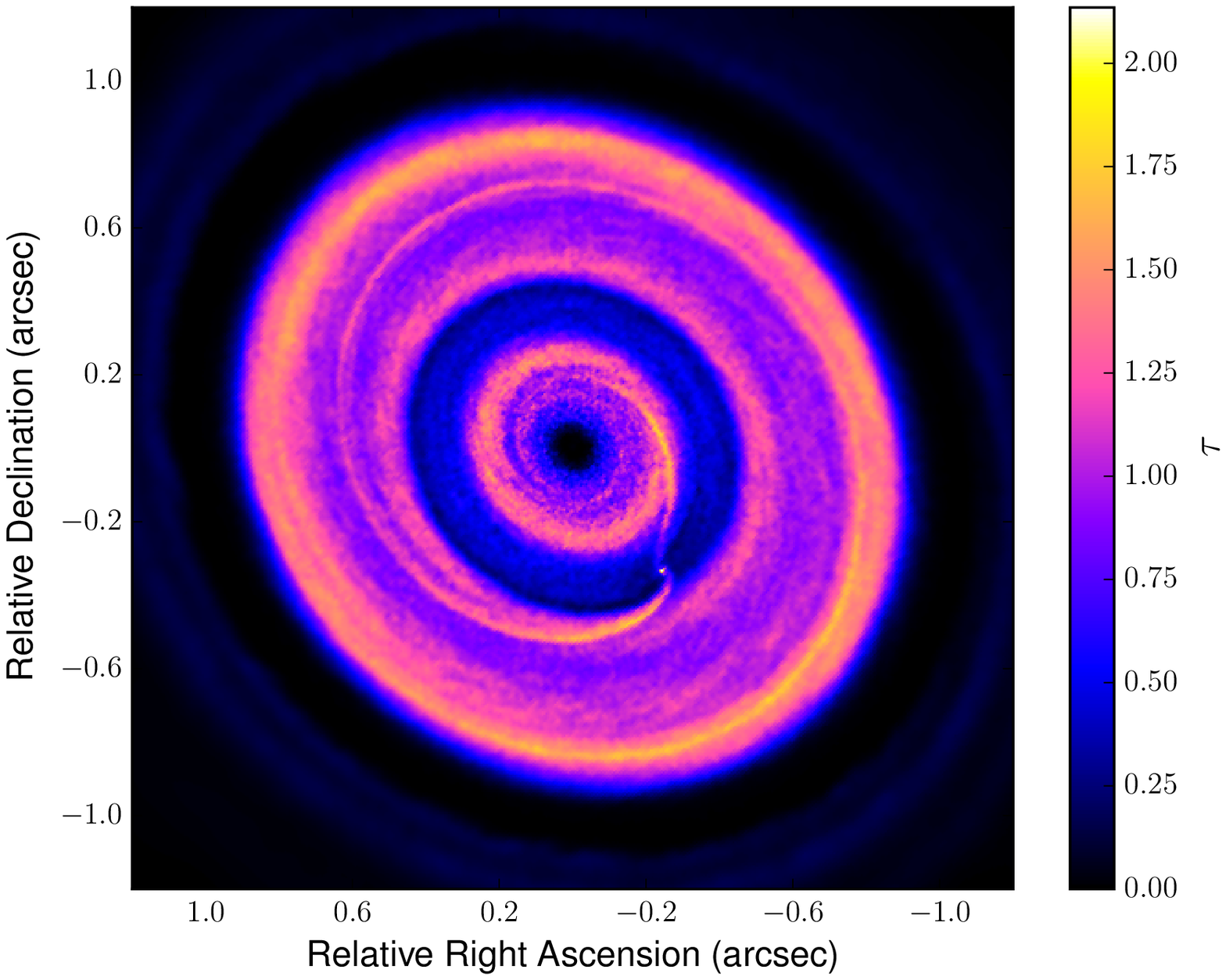}
\caption{Optical depth map at 1.3 mm of our model. Most of the disc is marginally optically thick, apart from the outer ring region.}
\label{fig:tausurf}
\end{center}
\end{figure}

We infer the width of the gap in the gas by computing the location of the  inner and outer gap edge. We  adopt the approach by \citet{dong16f}, which appears to work better for shallow gaps where the density does not drop below an empirical threshold (see also \citealt{kanagawa16a}). The gap outer and inner edge $r_{\mathrm{out,gap}}$ and $r_{\mathrm{in,gap}}$  are defined as the locations outside and inside the planet orbit where the dust surface density  reaches the geometric mean between its minimum value in the gap $\Sigma_{\mathrm{g}}(r_{\mathrm{min}})$ and its unperturbed value (i.e. computed from the initial gas density profile) at the same location $\Sigma_{\mathrm{g,0}}(r_{\mathrm{min}})$, i.e.
\begin{equation}
\Sigma_{\mathrm{g}}(r_{\mathrm{in,gap}})=\Sigma_{\mathrm{d}}(r_{\mathrm{out,gap}})\equiv \sqrt{ \Sigma_{\mathrm{g}}(r_{\mathrm{min}})  \times  \Sigma_{\mathrm{g,0}}(r_{\mathrm{min}})} .
\end{equation}
We infer a gap width of 27.7 au, which is consistent with the value of $\sim 5.8 H_{\mathrm{p}}$ found by \citet{dong16f}.
Fig.~\ref{fig:sigmagd} shows that the location of the minimum of the gas surface density is $\sim$ 4 au  from the location of the planet. This difference might be related to a combination of the initial non-uniform surface density profile and the fast planet migration in the last stages of the simulation (see Sect.~\ref{sect:planmigr}). The shape of the gas gap carved by a migrating and accreting planet has been investigated by \citet{dangelo08a}, finding a shape similar to that shown in our simulations (see Fig.~15 by \citealt{dangelo08a}).

Interestingly, in the low density outer disc regions, millimetre grains become less strongly coupled, producing a fast radial inward drift toward the pressure maximum at the outer gap edge. Since the aerodynamical coupling depends on the gas surface density, the non uniform gas distribution across the outer disc regions leads to a gradient in the dust radial velocities. This produces a flatter radial profile of the dust surface density that eventually changes the profile monotonicity, resulting in ring-like structures in the outer disc. The shape of the dust surface density in the outer disc region can be explained by the drift-dominated dynamics of non-growing grains with $\mathrm{St}< 1$ described by \citet{youdin02a} and \citet{birnstiel14a}. The bumps in the dust density in outer gap edge and in the outer disc produce a double-hump feature that will evolve to create an additional accumulation of dust at the outer edge of the gap. This confirms the analysis shown by \citet{birnstiel14a}, i.e. the radial drift of non-growing grains from the outer disc region produces a distinctive fingerprint in the dust surface density profile in the outer disc region.

It can be seen that the disc inside $\sim 15$ au has a significantly low mass. This occurs due to the resolution-dependent SPH numerical viscosity, which causes a high accretion rate onto the star \citep{lodato10a}.


\subsection{Comparison with ALMA observations}
\label{sect:simulaalma}

For the comparison of our model predictions with the results of the ALMA observations we adopted the procedure described in Section~\ref{sec:rf}.
Fig.~\ref{fig:cfrmodel} shows the comparison between the observed ALMA image and the simulated observations of our best model at the same wavelength. We note that the spiral structure observed in our hydrodynamic simulations (see Fig.~\ref{fig:renderingsigmagd}) is not observed by ALMA due to the limited angular resolution of the observations. The right panel in Fig.~\ref{fig:cfrmodel} shows the map of the data $-$ model residuals. The residuals map was obtained by first deriving the residuals for the real and imaginary parts of the visibility function on the same $\left(u,v\right)$ points as in the ALMA observations, and then imaging them, using the same procedure as for the imaging of the ALMA observed visibilities (Section~\ref{sec:almaobs}).

In Fig.~\ref{fig:cfrphotom} we show the data vs model comparison in terms of the azimuthally averaged surface brightness at different angular separations from the disc center. This plot was obtained by carrying out aperture photometry on both the ALMA map and the model synthetic image using the geometry parameters reported in Sect.~\ref{sect:almaobsconti} for the Elias 24 disc.

Apart from the bright central core, the surface brightness in our model provides a reasonable match to the gap and ring like structure observed in Elias 24. Starting from the radial profiles of the surface brightness, we can measure the discrepancies of our model with respect to the observed data, taking into account the error on each value of the surface brightness profile. The errorbars are calculated by dividing the rms noise of the observations by the square root of the number of independent beams in each annulus, in order to take into account the increase of the pixel number with the aperture extent.
In the outer disc (i.e. $\gtrsim 1''$ from the central star), the mean value of the absolute values of the residuals is $\left(7.3 \pm 1.9\right)$\% of the observed fluxes, while for the inner central core ($\lesssim 0.3''$) the difference is $\left(39.8 \pm 0.2\right)$\% of the observed fluxes. In the gap and outer disc edge region, i.e. in an annulus between 0.4$''$ and 1$''$ from the central star, the mean value of the absolute values of the residuals is $\left(5.1\pm 0.4\right)$\% of the observed fluxes. 
Importantly, we reproduce the change of concavity in the outer disc at $\sim$ 0.8$''$. This change of concavity can be explained by the differential radial motion of large dust grains from the outer radius (see Sect.~\ref{sect:resulstruct}). Beyond $\sim$ 1.15$''$ from the central star the relative difference between our model and the observation increases with distance from the central star. This mismatch might be related to a reduced dust density in the outer region due to the radial inflow of dust grains. Moreover, this slight mismatch might be related to the fact that the simplified two-grain-size treatment does not take into account the additional pile up of larger grains ($s_{\mathrm{grain}}>$ 1 mm) coming from the outer disc. Larger grains are expected to be more trapped at the pressure maximum, resulting in an additional small flux at the gap outer edge (see Sect.~\ref{sec:limits}).
 
The measured total flux of the disc from the simulated model map is $309.2 \pm1.1$ mJy, $\sim 16\%$ lower than the total flux of the real observation. However, most of the difference comes from the bright central core.
 Fig.~\ref{fig:tausurf} shows the map of the optical depth of our model. Most of the disc is marginally optically thick, while the ring in the outer region is characterized by a slightly higher optical depth $\tau\sim 2$, which is related to the pile-up of large grains spiralling inward from the outer disc (see Sect.~\ref{sect:resulstruct}).

\section{Discussion}
\label{sect:discussion}

\subsection{Gap opening}
\label{sect:gapopening}
\begin{figure}
\begin{center}
\includegraphics[width=0.48\textwidth]{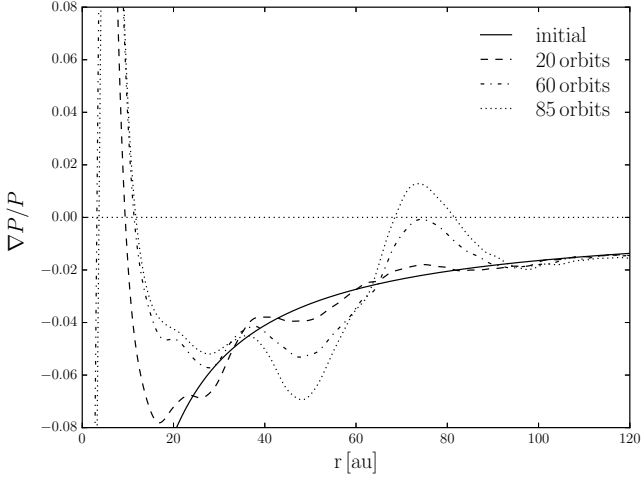}
\caption{Evolution of the azimuthally averaged profile of the pressure radial gradient. The solid line indicates the profile at the start of the simualation. Different lines correspond to different times along the simulations at which the planet mass reaches a value of (dashed) $M_{\mathrm{p, gap}}$ (see Eq.~\ref{eq:mlimgas}), (dotted-dashed) $2M_{\mathrm{p, gap}}$ and (dotted) $0.7\, \mathrm{M_{\mathrm{J}}}$.}
\label{fig:derivp}
\end{center}
\end{figure}

Our simulations show that, consistent with different numerical works \citep{rosotti16a,dipierro17a}, once the planet mass reaches a value of
\begin{equation}
M_{\mathrm{p, gap}}\approx 0.1\,M_{\mathrm{p,th}} \sim 0.19 \,\mathrm{M_{\mathrm{J}}},
\label{eq:mlimgas}
\end{equation}
where \citep{lin93a}
\begin{equation}
\frac{M_{\rm p,th}}{M_{\star}} \equiv 3 \left( \frac{H_{\mathrm p}}{r_{\mathrm p}} \right)^3 ,
\label{eq:cond1}
\end{equation}
the planet starts to weaken the pressure profile in its neighbourhood. Fig.~\ref{fig:derivp} shows the evolution of the radial profile of the pressure gradient along the simulation.  
The planet reaches a mass of $M_{\mathrm{p, gap}}$ after $\sim$ 20 orbits, 
 keeping carving the gap along the simulation and leading to a local reduction of the inward radial drift of dust grains from the outer disc region.
Once the planet reaches a mass of $\sim 0.45\,\mathrm{M_{\mathrm{J}}}$ (after $\sim$ 60 orbits), a pressure maximum at the gap outer edge is created, leading to a further accumulation of particles \citep[e.g.][]{lambrechts14a,dipierro16a}. This value for the minimum planet mass able to create a pressure maximum outside the planetary orbit is consistent with the expectations of the gap opening process limited by pressure forces $M_{\rm p} \sim 2M_{\mathrm{p, gap}} \approx 0.4\,\mathrm{M_{\mathrm{J}}}$. In detail, the minimum planet mass able to create a pressure maximum in the outer disc is expected to be $\sim\mathrm{max}\left( 2M_{\mathrm{p, gap}},M_{\mathrm{p,visc}}\right)$,
where
\begin{equation}
 \frac{M_{\rm p,visc}}{M_{\star}}\equiv\left(\frac{27 \pi}{8}\right)^{1/2} \left(\frac{H_{\mathrm p}}{r_{\mathrm p}}\right)^{5/2}\alpha_{\mathrm{SS}}^{1/2} 
\label{eq:cond2}
\end{equation}
is the limit planet mass under which the viscous refilling of the gap is faster than the gap opening \citep{armitage10a}. In our model, due to the relatively high thickness of the disc (the height at the planet position is $H_{\mathrm p}/r_{\mathrm p}\sim 0.08$), the pressure forces are the dominant ones trying to close the gap, i.e. $\mathrm{max}\left( 2M_{\mathrm{p, gap}},M_{\mathrm{p,visc}}\right)=2M_{\mathrm{p, gap}}$.
We stress that there is no universally accepted criterion for the gap-opening planet mass. The widely used gap-opening criterion formulated by \citet{crida06a} gives a mass of $2.8 \,\mathrm{M_{\mathrm{J}}}$, which is expected to correspond to the case when planet's co-orbital gas surface density drops to 10\% its initial value.

\subsection{Planet migration}
\label{sect:planmigr}
The planet migrates from a radial distance of 65 to 60.4 au in 100 orbits (measured at 65 au, the initial planet location) towards the central star. Since the gas gap carved by the planet is shallow, it is reasonable to expect that the planet migration is described by the Type I regime \citep{kley12a}.
Assuming a gas disc model with a power law surface density and temperature radial profile, i.e. $\Sigma_{\mathrm{g}}\propto r^{-p}$ and $T\propto r^{-q}$, the total tidal torque, given by the contribution of the Lindblad and the unsaturated static linear corotation torque (Eqs.~14 and 17 of \citealt{paardekooper10a} or Eqs.~ 3, 6 and 7 of  \citealt{paardekooper11a}), is given by 
\begin{equation}
\Gamma=-\gamma\Gamma_{0} ,
\label{eq:const}
\end{equation}
where
\begin{equation}
\gamma=2.5+0.9\,q-0.1\,p-0.7\left(\frac{3}{2}-p\right)
\end{equation}
and
\begin{equation}
 \Gamma_{0} =\left (\frac{M_{\mathrm{p}}}{M_{\star}} \right )^2 \left(\frac{H_{\mathrm{p}}}{r_{\mathrm{p}}} \right)^{-2}\Sigma_{\mathrm{p}}r_{\mathrm{p}}^4\Omega_{\mathrm{p}}^2,
 \label{eq:torque}
\end{equation}
where $\Sigma_{\mathrm{p}}$ and $\Omega_{\mathrm{p}}$ denote the gas surface density and Keplerian angular frequency at the planet location $r_{\mathrm{p}}$, respectively. The formula expressed in Eq.~\ref{eq:torque} has been tested against 2D and 3D numerical hydrodynamical simulations of locally isothermal discs \citep{paardekooper10a,dangelo10b}, finding good agreement in viscous discs, where non-linear effects are strongly damped (for more details see \citealt{kley12a} and \citealt{baruteau14a}). 
The migration timescale is simply given by 
\begin{equation}
t_{\mathrm{migr,I}}=\frac{r_{\mathrm{p}}}{\left |\dot{r}_{\mathrm{p}} \right |}=\frac{J_{\mathrm{p}}}{2\left |\Gamma \right |}=\frac{1}{2\gamma} \left(\frac{H_{\mathrm{p}}}{r_{\mathrm{p}}} \right)^{2} \frac{1}{M_{\mathrm{p}}\Omega_{\mathrm{p}}}\frac{M_{\star}^2}{\Sigma_{\mathrm{p}}r_{\mathrm{p}}^2}
,
\label{eq:tmigrI}
\end{equation}
where $J_{\mathrm{p}}$ is the angular momentum of the planet, i.e. $M_{\mathrm{p}} r_{\mathrm{p}}^2\Omega_{\mathrm{p}}$. 
We measure the migration timescale in our simulations 
and compare it with the theoretical estimates, by taking into account both the decrease of the gas density at the planet location and the increase of the planet mass due to accretion onto the planet (see Sect.~\ref{sect:planetaccr}). 
To avoid numerical artifacts related to the initial conditions, we start measuring the migration timescale after 10 planetary orbits. 
\begin{figure}
\begin{center}
\includegraphics[width=0.48\textwidth]{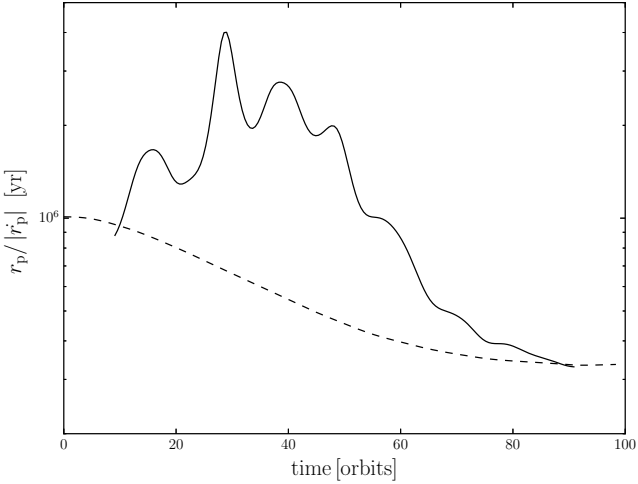}
\caption{Comparison between (solid) the migration timescale of the planet in our model and (dashed) the expected Type I migration timescale.}
\label{fig:tmigr}
\end{center}
\end{figure}
\newtext{
Fig.~\ref{fig:tmigr} compares the migration timescale of the planet predicted by our model and the typical Type I migration timescale (Eq.~\ref{eq:tmigrI}). Shortly after the start of the simulations, the migration timescale increases and oscillates around a value that decreases with time. The initial trend might be ascribed to the strong time evolution of the non-linear corotation torque (not included in Eq.~\ref{eq:torque}, \citealt{paardekooper10a,paardekooper11a}). In detail, at the beginning of the simulation, the initial gradient in specific vorticity and entropy generates asymmetries in the torques at the two horseshoe U-turns close to the planet, producing a strong positive corotation torque which leads to a deceleration of the classical Type I inward migration. 
Afterwards, the oscillation of the migration timescale might be related to the oscillation of the corotation torque induced by the flattening of the original vorticity and entropy gradient, a phenomenon mostly referred to as saturation of the corotation torque \citep{masset01a,paardekooper11a}. 
The oscillation period is roughly equal to the libration timescale of the material near the outer edges of the horseshoe region, which is around 15 orbits. 
Importantly, by comparing the expected libration and viscous timescales across the coorbital zone, we find that, after the planet reaches $\sim 0.2 \,\mathrm{M_{\mathrm{J}}}$ (after $\sim$ 25 orbits), the libration timescale becomes lower than the viscous timescale, implying that viscosity is not efficient enough to re-estabilish the vorticity gradient and prevent saturation \citep{masset02a}. 
The oscillations on a typical libration time observed are similar to the ones found for non-accreting planets (see Fig.~2 by \citealt{paardekooper10a}). The only noticeable difference is the decrease of the quantity $r_{\rm p} / \vert \dot{r _{\rm p}}\vert$, which might be related to the increase of the planet mass sustained by accretion (see Eq.~\ref{eq:tmigrI} and Sect.~\ref{sect:planetaccr}).}
Importantly, after a few oscillations the migration time approaches a stationary state, close to the nominal Type I migration expressed in Eq.~\ref{eq:tmigrI}.
Moreover, we do not see any sharp transition to the slower Type II migration, probably due to strong  feedback from the coorbital mass in the horseshoe region \citep{masset03a}, which is around 60\% of the initial value at the end of the simulation. 
However, an accurate analysis of the tidal interaction between the disc and an accreting and migrating planet is beyond the scope of the paper, due to the complexity involved \citep[e.g.][]{paardekooper06a,kley09a,paardekooper14a,benitez15a,lega15a}.


\subsection{Planet accretion}
\label{sect:planetaccr}
The planet accretes from the surrounding gas and dust disc, growing along the simulation from $0.15\, \mathrm{M_{\mathrm{J}}}$ to $\sim 0.9 \, \mathrm{M_{\mathrm{J}}}$ in 100 orbits (at 65 au). 
Planet accretion can be simply modelled by a sequence of stages related to the efficiency of pebble and gas accretion \citep[e.g.][]{helled14a}.
Concerning pebble accretion, when the planet affects the local gas pressure profile or the tidal torque is high enough to affect the inward drift of uncoupled grains \citep{dipierro17a}, the accretion of pebbles is expected to be strongly reduced \citep{lambrechts12a,bitsch15a}. This mass, mostly known as isolation mass, is given by (assuming $\epsilon \ll 1$) 
\begin{equation}
\left( \frac{M_{\rm p}}{M_{\rm \star}}\right)_{\mathrm{iso}} \approx \mathrm{min} \left[\frac{M_{\mathrm{p, gap}}}{M_{\rm \star}}, 1.38 \left(  \frac{\zeta}{ \mathrm{St}} \right)^{3/2}  \left(\frac{H_{\mathrm{p}}}{r_{\rm p}}\right)^3\right],
\label{eq:sum2}
\end{equation}
where
\begin{equation}
\zeta \equiv -\left. \frac{\partial \log P}{\partial \log r} \right|_{r_{\rm p}} 
\label{eq:zeta}
\end{equation}
is the exponent that characterises the steepness of the radial pressure profile of the disc. 
The second term in Eq.~\ref{eq:sum2} is related to the halting of uncoupled dust grains by the balancing effect of the tidal and drag torque outside the planetary orbit \citep{dipierro17a}. Due to the low Stokes numbers of pebbles, $M_{\mathrm{p, iso}} = M_{\mathrm{p, gap}}$. 
Therefore, since after the first $\sim$ 20 orbits the mass of the planet reaches $M_{\mathrm{p, iso}}$, most of the accreting flow is in the gas phase. 
Moreover, since for most of the simulation $M_{\mathrm{p}}\gtrsim \sqrt{2} M_{\mathrm{p, iso}}$, it is reasonably expected that the planet accretion is in the runaway phase \citep{pollack96a,helled14a}. 
\newtext{
In detail, while for low mass cores the accretion is limited by the cooling rate to evacuate the energy of the accreting flow  \citep{hubickyj05a,ayliffe09a}, more massive cores  essentially accrete at the runaway rate where gravity dominates the effects of thermal support. In our locally isothermal simulations, the extra heat generated by compression of the collapsing over-density is instantaneously radiated away and the accretion flow of the gas on the planet is set by the properties of the disc in the vicinity of the planet. 
}

 We can compare the mass accretion rate measured in the simulation with the expectation of the typical disc-limited accretion rate given by the flux of gas in the planet's cross section. The expected accretion rate is given by \citep{dangelo08a}
\begin{equation}
\dot{M}_{\mathrm{p}} = 
\begin{cases}
\displaystyle C_{\mathrm{B}}  \Omega_{\mathrm{p}}\Sigma_{\mathrm{p}} r_{\mathrm{p}}^2 \left(\frac{r_{\mathrm{p}}}{H_{\mathrm{p}}} \right)^7 \left(\frac{M_{\rm p}}{M_{\star}}\right)^{3} , & \text{$M_{\mathrm{p}} < M_{\mathrm{tr}}$,} \\[0.5cm]
\displaystyle \frac{C_{\mathrm{H}}}{3} \Omega_{\mathrm{p}}\Sigma_{\mathrm{p}}r_{\mathrm{p}}^2 \left(\frac{r_{\mathrm{p}}}{H_{\mathrm{p}}} \right) \left(\frac{M_{\rm p}}{M_{\star}}\right), & \text{$M_{\mathrm{p}} \geq M_{\mathrm{tr}}$,}
\end{cases}
\label{eq:def_accre}
\end{equation}
where $C_{\mathrm{B}}$ and $C_{\mathrm{H}}$ are dimensionless coefficients of order unity, and
\begin{equation}
M_{\mathrm{tr}}=\frac{M_{\star}}{\sqrt{3}}\sqrt{\frac{C_{\mathrm{H}}}{C_{\mathrm{B}}}} \left(\frac{H_{\mathrm{p}}}{r_{\mathrm{p}}} \right)^3\sim 0.22 \,\mathrm{M_{\mathrm{J}}}
\label{eq:masstrans}
\end{equation}
indicates the mass transition between the Bondi- and Hill-type gas accretion regimes, differing from each other by the characteristic planet's accretion radii (for more details see \citealt{dangelo10a}).
\begin{figure}
\begin{center}
\includegraphics[width=0.48\textwidth]{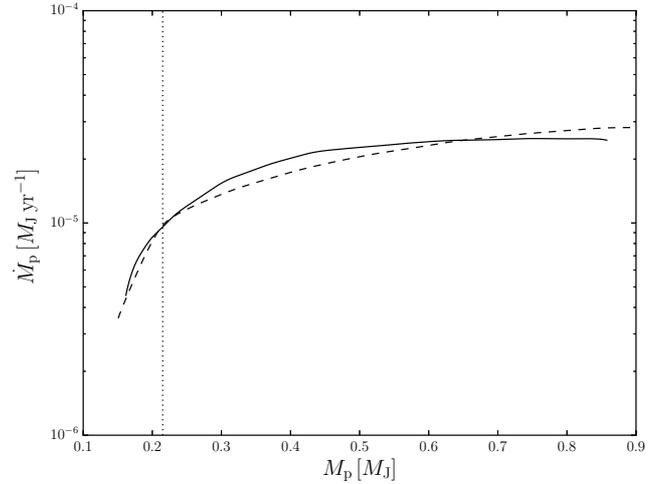}
\caption{Comparison between (solid) the accretion rate of the planet in our model and (dashed) the expected runaway accretion rate expressed in Eq.~\ref{eq:def_accre}. The vertical line indicates the transition planet mass between the Bondi- and Hill-type gas accretion regimes (Eq.~\ref{eq:masstrans}).}
\label{fig:acc}
\end{center}
\end{figure}

Fig.~\ref{fig:acc} shows the accretion rate of the planet along the simulation.
The accretion rate is measured as the rate at which gas and dust particles pass into the radius of the sink, which is 0.25 au, $\sim$ 9 (5) and 16 (28) times smaller than the Hill (Bondi) radius\footnote{\citet{lissauer09a} found that the gas gets accreted onto the planet if it flows through a sphere around the planet of radius $\sim 0.25 R_{\mathrm{H}}$.} at the beginning and at the end of the simulation, respectively. This choice for the value of the sink radius ensures a good level of physical accuracy \citep{ayliffe09a,machida10a}.
Fig.~\ref{fig:acc} shows that the accretion rate computed in our simulations is very close to the expectation of the disc-limited accretion rate. We have used the same values of $C_{\mathrm{B}}$ and $C_{\mathrm{H}}$ found by \citet{dangelo08a} scaled by a factor of $\sim 2$ (i.e., we use $C_{\mathrm{B}}= 5.13$ and $C_{\mathrm{H}} = 1.75$), given by the different amount of physical viscosity adopted in this work \citep{szulagyi14a}. 
In the initial part of the simulation, the accretion rate increases with the planet mass until it reaches a nearly constant value of  $2.5\, \times 10^{-5} \,\mathrm{\mathrm{M_{\mathrm{J}}}\, yr^{-1}}$. The trend of the accretion rate  with planet mass seen in Fig.~\ref{fig:acc} is consistent with the typical one of the disc-limited accretion rate close to the transition between the Bondi- and the Hill-type accretion rate (see Eq.~\ref{eq:masstrans}). The transition mass (see Eq.~\ref{eq:masstrans}), corresponding to the mass at which the Hill radius becomes smaller than the Bondi radius (see Fig.~4 by \citealt{dangelo08a}), occurs when the mass reaches $\sim 0.22 \,\mathrm{M_{\mathrm{J}}}$ (after $\sim$ 27 planetary orbits).
Our simulation does not show any decreasing planetary accretion rate in the late stages, since the gas gap is very shallow at the end of the simulation, so that the surrounding disc is still able to supply gas to the planet's vicinity.

\subsection{The fate of the planet}
\label{sect:fatepla}
In order to predict if the planet will survive to its migration, we can compute the reduction of its semi-major axis after the end of the simulation. Let us consider a planet migrating from a distance of $r_{\mathrm{p,i}}\sim60.4$ au from the central star and growing from the mass $M_{\mathrm{p,i}}\sim0.9 \, \mathrm{M_{\mathrm{J}}}$ to a mass $M_{\mathrm{p,f}}$ given by the gap-opening mass, when the planet migration is expected to be significantly slowed down \citep{kley12a}. We assume that the gap-opening mass is given by the \citet{crida06a}'s criterion, corresponding to a drop of the local surface density to a factor $\sim 10\%$ of the initial value, i.e. \citep{baruteau14a}
\begin{equation}
\frac{M_{\mathrm{p,f}}}{M_{\rm \star}}=100 \alpha_{\mathrm{SS}} \left(\frac{H_{\mathrm{p}}}{r_{\mathrm{p}}}\right)^2 \left [ \left(X+1\right)^{1/3} - \left(X-1\right)^{1/3} \right]^{-3}  ,
\label{eq:condcrida}
\end{equation}
with
\begin{equation}
X=\sqrt{1+\frac{3}{800\alpha_{\mathrm{SS}}} \frac{H_{\mathrm{p}}}{r_{\mathrm{p}}}}.
\end{equation}
The gap opening mass is $\sim\, 3.4 \,\mathrm{M_{\mathrm{J}}}$.
Assuming that the planet is in Type I migration following $\dot{r}_{\mathrm{p}}=-r_{\mathrm{p}}/t_{\mathrm{migr}}$ with $t_{\mathrm{migr}}$ expressed in Eq.~\ref{eq:tmigrI} and accreting in runaway mode following the second expression of Eq.~\ref{eq:def_accre} , the reduction of its radial location is given by
\begin{equation}
\int_{r_{\mathrm{p,f}}}^{r_{\mathrm{p,f}}} \frac{d r_{\mathrm{p}}}{r_{\mathrm{p}}}= -\int_{M_{\mathrm{p,f}}}^{M_{\mathrm{p,f}}} \xi \frac{d M_{\mathrm{p}}}{M_{\star}},
\label{eq:integmigr}
\end{equation}
where $\xi$ is ratio between the typical growth timescale $t_{\mathrm{growth}}=M_{\star}/\dot{M}_{\mathrm{p}}$ and migration timescale $t_{\mathrm{migr}}$, i.e.
\begin{equation}
\xi=\frac{t_{\mathrm{growth}}}{t_{\mathrm{migr,I}}}=\frac{3\gamma}{C_{\mathrm{H}}}\left(\frac{H_{\mathrm{p}}}{r_{\mathrm{p}}} \right)^{-1}.
\end{equation}
Since the disc aspect ratio has a weak dependence on $r_{\mathrm{p}}$, i.e. $H_{\mathrm{p}}/r_{\mathrm{p}}\propto r_{\mathrm{p}}^{0.03}$ (see Sect.~\ref{sect:initcond}), we assume that $\xi$ is constant with $r_{\mathrm{p}}$.
The integration expressed in Eq.~\ref{eq:integmigr} yields
\begin{equation}
r_{\mathrm{p,f}}=r_{\mathrm{p,i}} \exp{\left[- \xi \left(\frac{M_{\mathrm{p,f}}-M_{\mathrm{p,i}}}{M_{\star}}\right)\right]}\approx 53 \,\mathrm{au} .
\label{eq:rpf}
\end{equation}
It is worth remarking that, before the planet reaches the mass $M_{\mathrm{p,f}}$, it is expected that the gradual depletion of the coorbital mass leads to a reduction of the planet accretion rate and migration (see Figs.~4 and 6 by \citealt{dangelo08a}). Therefore, the value given in Eq.~\ref{eq:rpf} can be considered an underestimation of the final radial location of the planet.
Once a deep gap is carved, assuming that most of the angular momentum content is in the gas rather than in the planet, the viscous evolution of the disc drives the planet migration \citep{baruteau14a}. In this Type II regime, the migration timescale is given by \citep{ivanov99a} 
\begin{equation}
t_{\mathrm{migr,II}}=\frac{M_{\mathrm{p}}+M_{\mathrm{g}} \left( r_{\mathrm{p}}\right)}{M_{\mathrm{g}}\left ( r_{\mathrm{p}}\right)} t_{\rm{visc}} \left( r_{\mathrm{p}}\right),
\end{equation}
where 
\begin{equation}
t_{\rm{visc}}\left( r_{\mathrm{p}}\right)=\frac{2}{3}\frac{r_{\mathrm{p}}^2}{\nu \left (r_{\mathrm{p}} \right)}
\end{equation}
is the viscous timescale and $M_{\mathrm{g}}\left( r_{\mathrm{p}}\right)=4\pi r_{\mathrm{p}}^2\Sigma_{\mathrm{p}}$ is a measure of the local gas disc mass. 
Assuming a planet with mass equal to the gap-opening mass (Eq.~\ref{eq:condcrida}), the Type II migration timescale is $\sim \, 3.2$ Myr.
This result confirms the analysis shown by \citet{crida17a}, i.e. a planet in the runaway accretion embedded in thick discs is saved from Type I migration without being lost into its host star. 

\begin{figure*}
\begin{center}
\includegraphics[height=0.293\textwidth,trim={0 0 0. 0}]{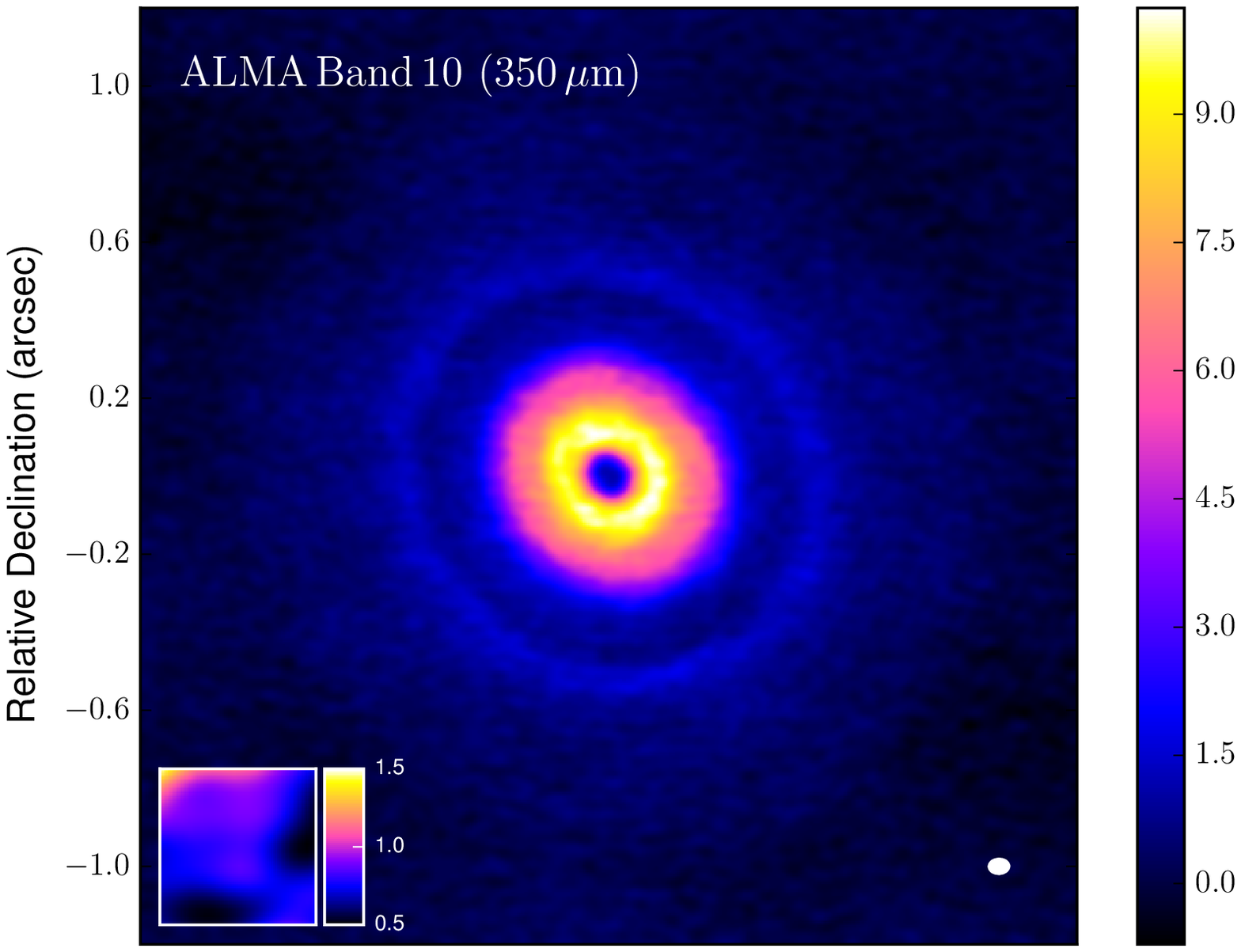}
\includegraphics[height=0.293\textwidth,trim={1.65cm 0 0. 0},clip]{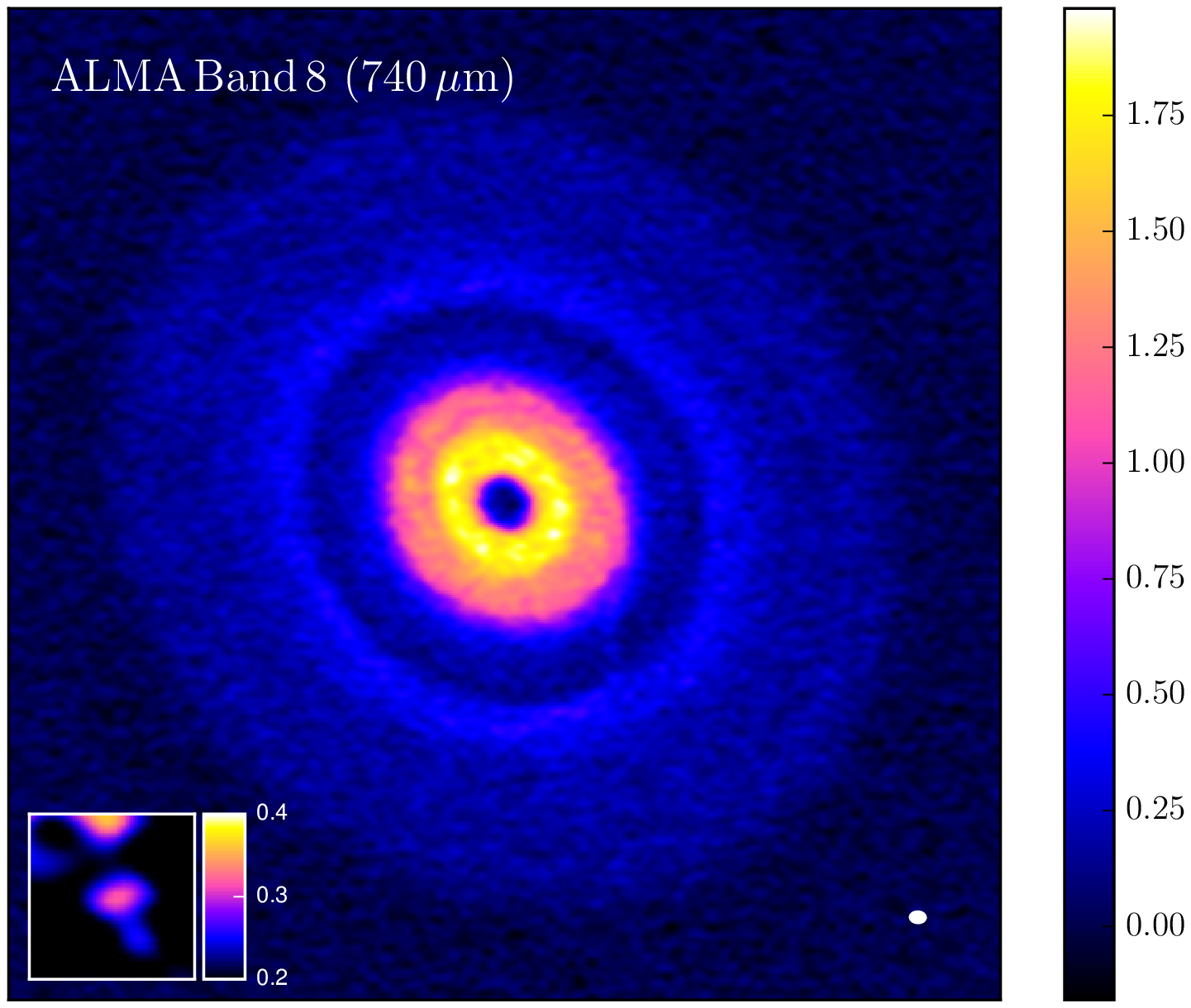}
\includegraphics[height=0.293\textwidth,trim={1.65cm 0 0. 0},clip]{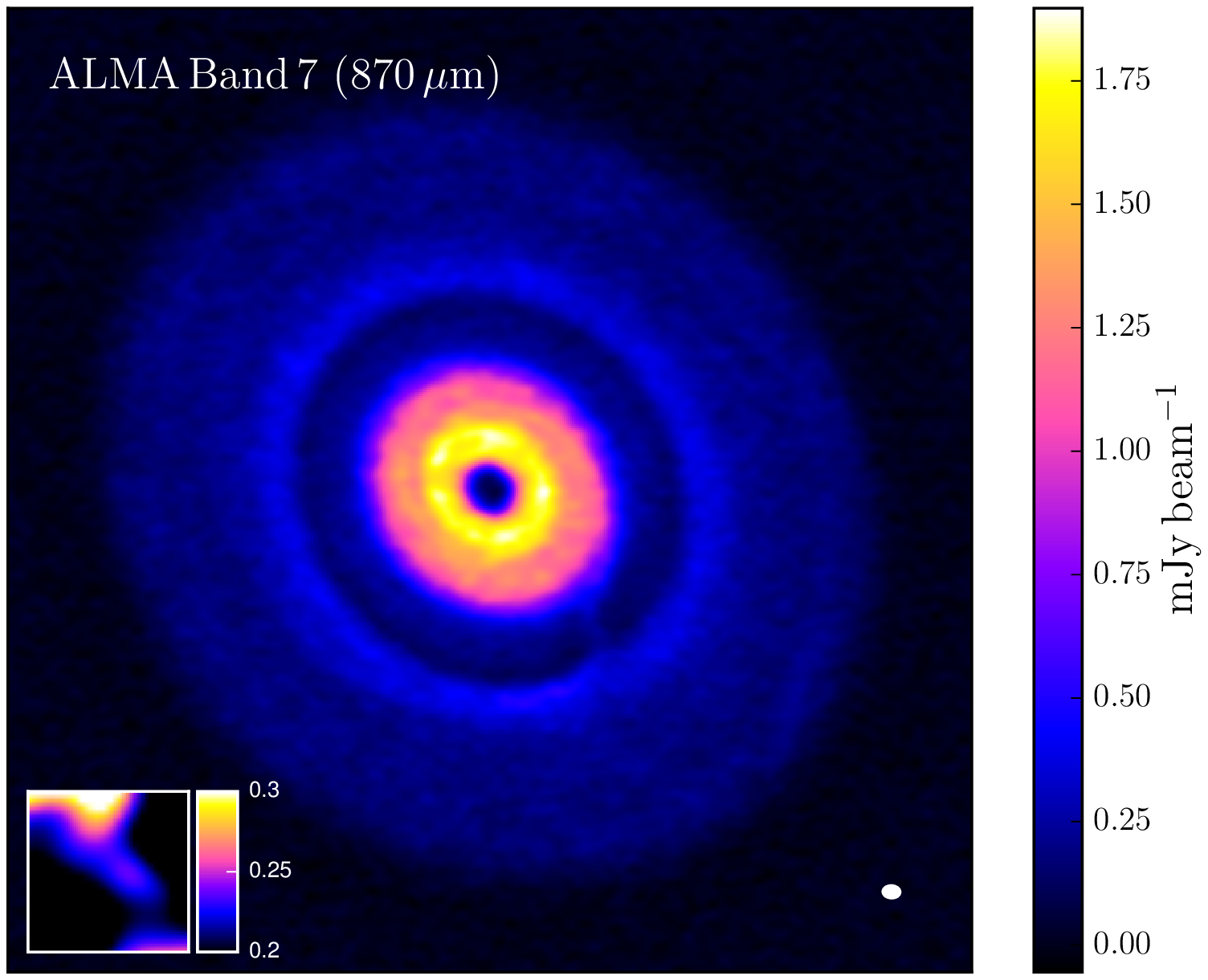}
\centerline{\includegraphics[height=0.293\textwidth,trim={0 0 0. 0.cm}]{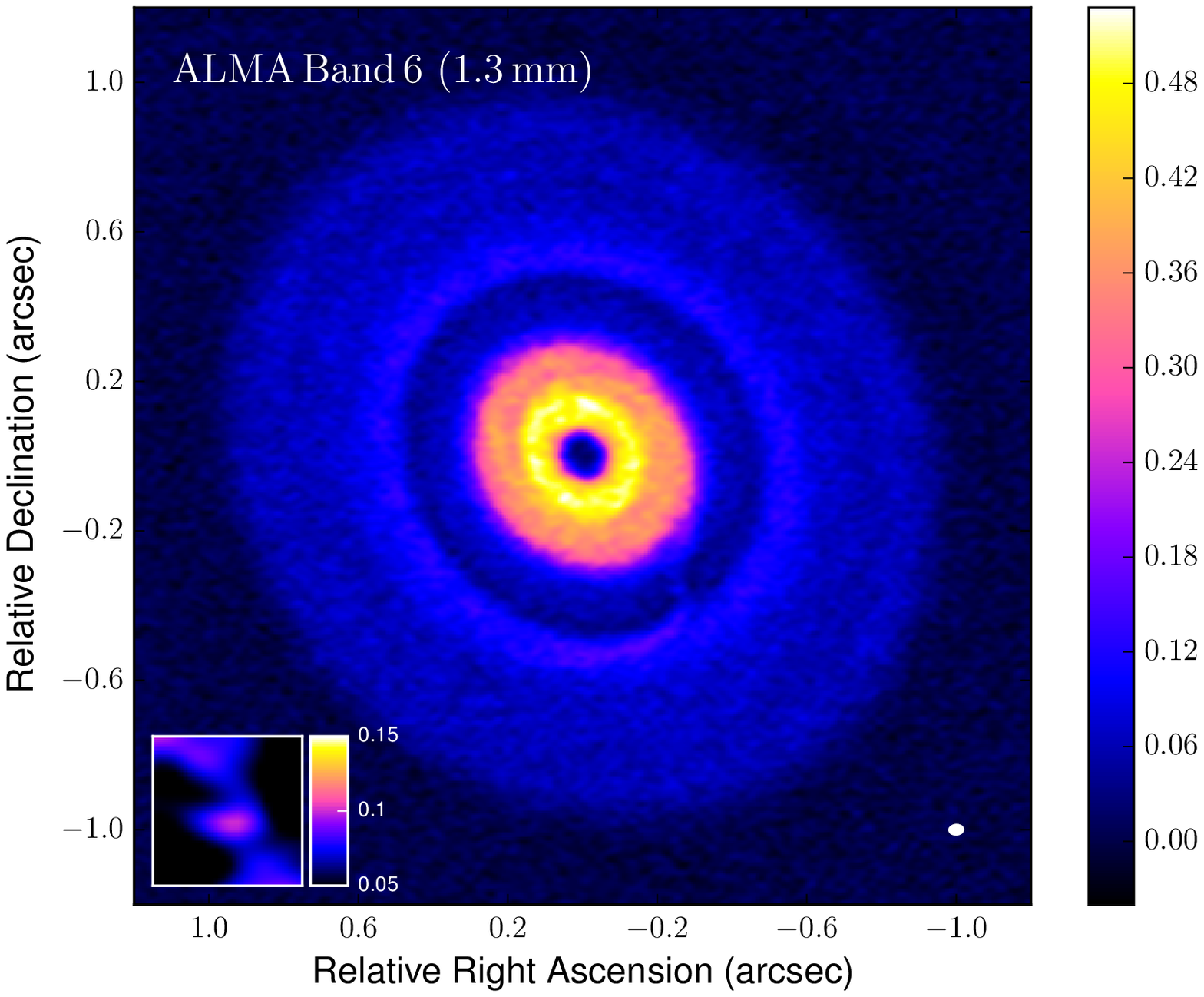}
\includegraphics[height=0.293\textwidth,trim={1.65cm 0 0. 0.cm},clip]{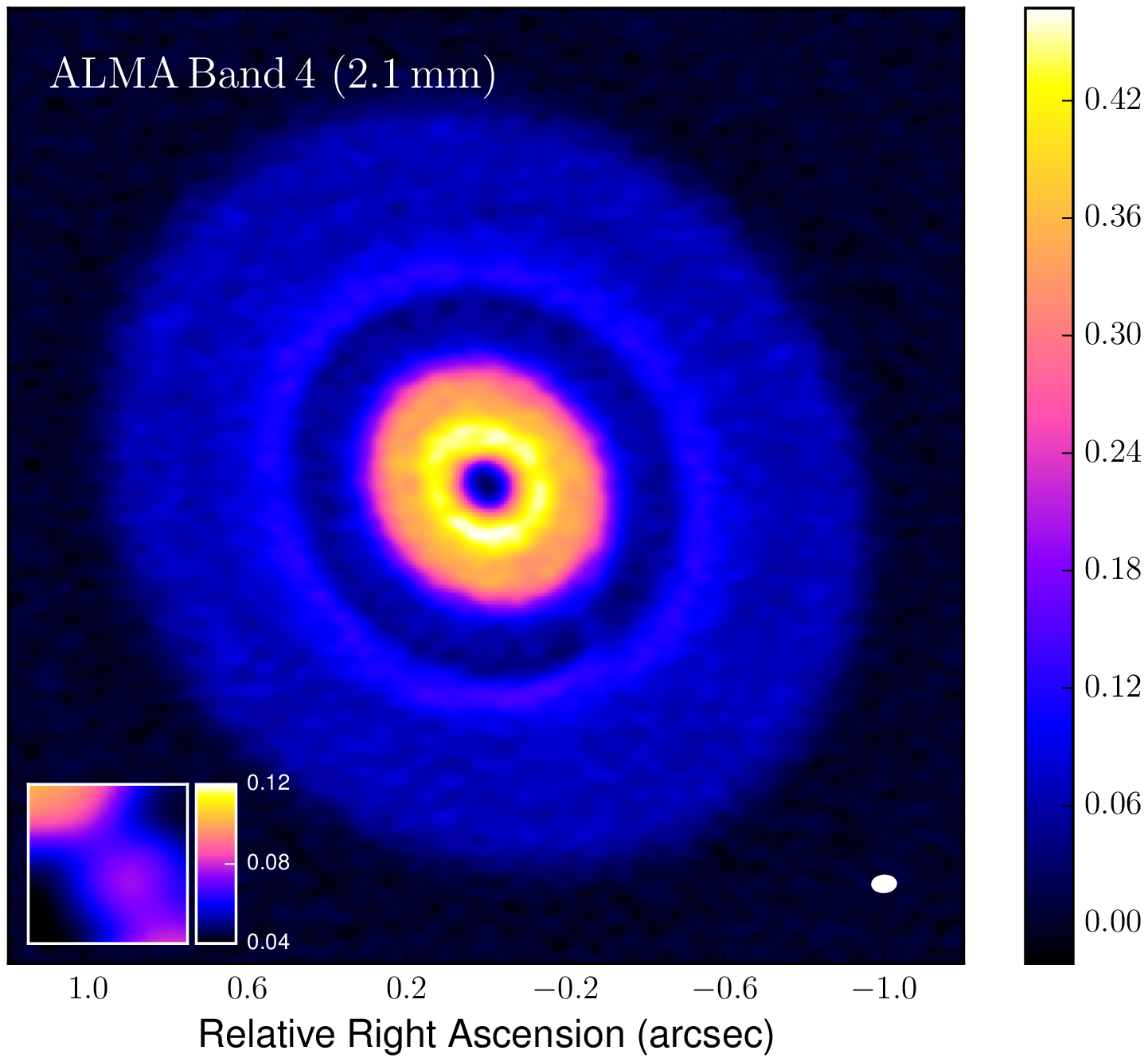}
\includegraphics[height=0.293\textwidth,trim={1.65cm 0 0. 0.cm},clip]{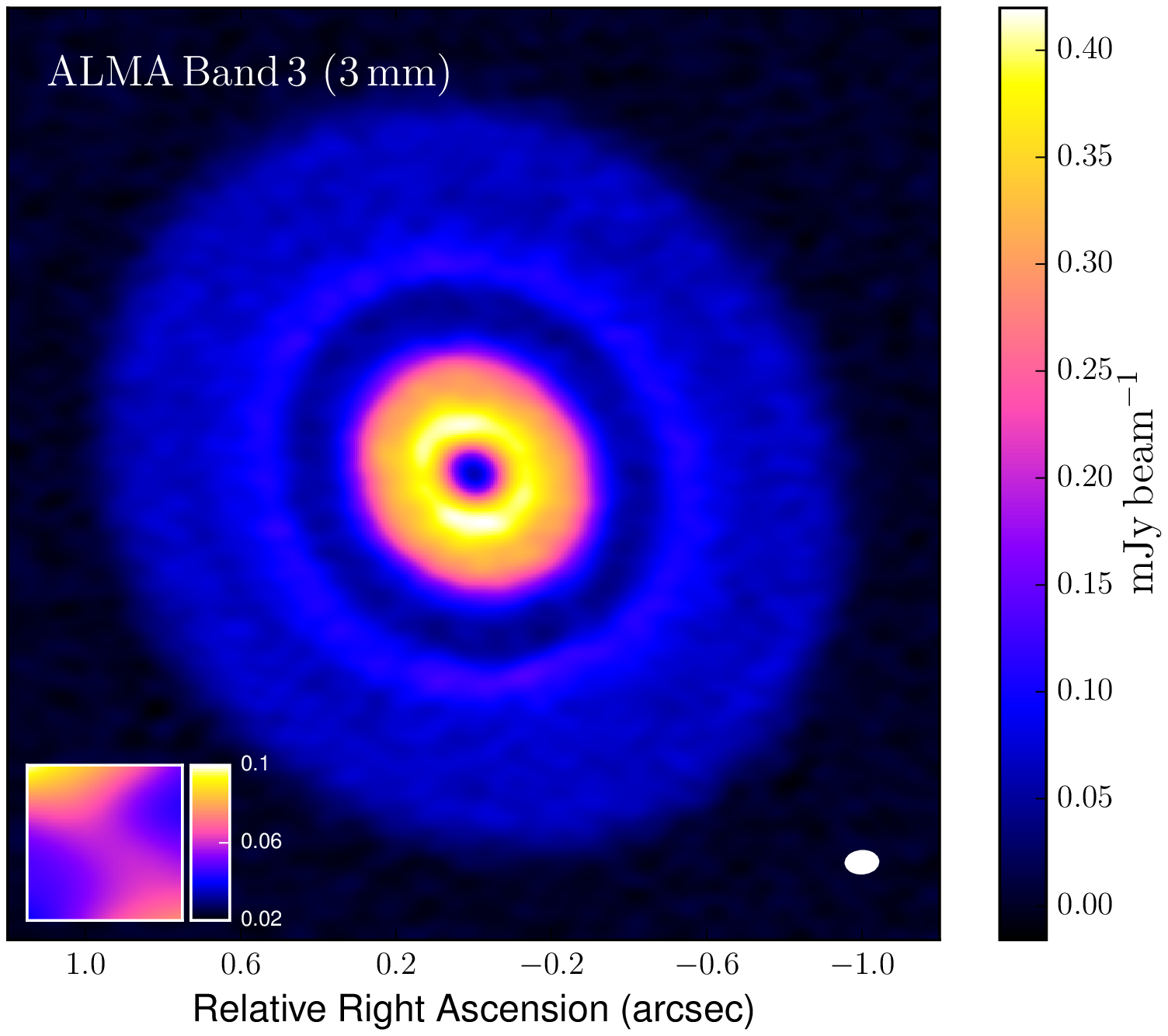}}
\caption{Comparison among synthetic observations in different ALMA Bands of our model of Elias 24. A zoom-in view of the circumplanetary disc is shown in the lower left corner of each panel. The white ellipse in the lower right corners indicates the size of the synthesized beam: (top-left) 0.05'' $\times$ 0.037'', $\mathrm{P.A.} = -88^{\circ}$, (top-center) 0.036'' $\times$ 0.026'', $\mathrm{P.A.} = 87^{\circ}$, (top-right) 0.042'' $\times$ 0.03'', $\mathrm{P.A.} = 87^{\circ}$, (bottom-left) 0.035'' $\times$ 0.024'', $\mathrm{P.A.} = -86^{\circ}$, (bottom-center) 0.057'' $\times$ 0.039'', $\mathrm{P.A.} = -86^{\circ}$ and (bottom-right) 0.082'' $\times$ 0.055'', $\mathrm{P.A.} = -86^{\circ}$.}
\label{fig:cfrmodelbnads}
\end{center}
\end{figure*}

\subsection{The origin of the gap structure}
\label{sect:models}

\newtext{
Gap like features can be explained by a variety of phenomena occurring in protoplanetary discs, each of which requires particular conditions on the gas structure, optical, physical and chemical properties of dust grains and, in some cases, on the structure of the magnetic field and the ionization of the disc. 
Among all the models proposed to explain gaps in protoplanetary disc, the planetary hypothesis is the most important to test and eventually rule out since it may put strong constrains on formation timescales and eventually explain characteristics of observed exoplanets \citep{simbulan17a}. Moreover, the planetary hypothesis has been invoked in most of the recent investigations on substructures in protoplanetary discs \citep[e.g.][]{andrews16a,fedele17b,fedele17a,ragusa17a}.  
Assuming that the features observed in Elias 24 are due to an embedded protoplanet, our modelling shows a reasonable match with the ALMA observations, providing strict constraints on the gas structure of the disc. 
In detail, since the dust dynamics are heavily dependent on the gas surface density (see Eq.~\ref{eq:ts}), different choices for the steepness of the gas surface density or a different value of the gas mass would produce a different efficiency of the radial dust migration, leading to a different profile of the radial intensity in the gap and outer disc region. 
Regarding the planet mass, our simulations shows that the morphology of the gap induced by the planet, although hampered by the low resolution of the observation, is consistent with the gap shape produced  by a low mass planet able to marginally perturb the local pressure profile without creating a deep gap. 
Importantly, while most of the modelling studies in recent years assume that the planet properties (i.e. mass and radial position) are fixed along the simulations, we let the planet to migrate and accrete, providing further constraints on the gas disc model.  Since the migration and accretion rate depend on the planet mass and local gas surface density (see Eqs.~\ref{eq:tmigrI} and \ref{eq:def_accre}), the adoption of a particular gas surface density profile and gas mass provides strict constraints on the mass of the planet able to affect the local pressure profile and subsequently produce the surface density bump at the outer gap edge. Moreover, as described in Sect.~\ref{sect:fatepla}, if we run the simulation for longer, the gas and dust density structures reach a steady state when the planet carves a deep gap, leading to a strong reduction of its migration and accretion \citep{dangelo10a}. In this case, we cannot reproduce the shallow gap structure observed by ALMA. 
Future high resolution and high fidelity observations over a large range of wavelengths will allow testing of this hypothesis by detecting the emission from the circumplanetary disc in the gap (see Sect.~\ref{sect:deteccircum}, \citealt{szulagyi17a,zhu17a}).

Among the alternative scenarios for the origins of gaps in protoplanetary discs, a frequently invoked model is related to the presence of ice snow lines of molecular species. The location of the gap in Elias 24 might
suggest that the latter might be due to the existence of a CO snow
line at the location of the gap \citep{zhang15a,pinilla17a}. Recently,
\citet{stammler17a} investigated how the CO ice line
can affect the particle growth and dynamics, finding a
depletion in the dust density structure of millimetre grains at the
ice line. Although the \citet{stammler17a} models can produce gaps in $\sim$ mm-sized grains with a depth comparable with the one constrained for the Elias 24 disc, the predicted radial widths are of the order of 5-10 au, therefore narrower than the Elias 24 case (\emph{Stammler, priv. comm.}).
Further modelling of the evolution of dust particles in the presence of the CO snow line applied to the physical conditions of the Elias 24 disc is necessary to test the availability of this effect to explain the Elias 24 gap.

The gap can also be explained by a radial variation of ionization, that in turn generates a gradient of the angular momentum transport induced by magnetorotational instability \citep{flock15a}. This creates a gas pressure maxima at the edge of this so-called ``dead zone'', leading to a strong accumulation of large dust grains \citep{pinilla16a}.
}


\subsection{Detectability of the circumplanetary disc}
\label{sect:deteccircum}
One of the recent active topics of research in the field of protoplanetary discs is the direct detectability of forming planets and their associated circumplanetary discs still embedded in their parent protoplanetary discs \citep{szulagyi17a,zhu17a}.  Since the first detection in the LkCa15 system \citep{kraus12a,sallum15a}, inconclusive evidence for circumplanetary discs has been found in a number of sources using high-contrast imaging techniques at mid infrared wavelengths, where these discs are expected to emit most \citep{reggiani14a,quanz15a}. Moreover, the first candidates for circumplanetary discs have been detected by ALMA around a free floating planetary mass object OTS 44 \citep{bayo17a} and inside the disc cavity in HD 142527 \citep{boehler17a}.

As recently found by \citet{zhu17a} and \citet{szulagyi17a}, the detection of circumplanetary discs requires deep observations at extremely high resolution. In the case of the ALMA image of Elias 24 presented here, the detection of a disc around the planet is hampered by the low resolution of the ALMA observation. As can be observed in Fig.~\ref{fig:continuum} and \ref{fig:continuum_phot}, even the gap is only marginally resolved by our observations and any potential circumplanetary disc emission would be strongly contaminated by the background circumstellar disc flux. 
However, we can use our models to estimate the detectability of the circumplanetary disc in the gap by computing synthetic ALMA observations with higher resolution and sensitivity. The resolution needed to detect the circumplanetary disc is of the order of the size of the disc, i.e. $\sim$ 0.025'', consistent with the typical disc diameter $\sim R_{\mathrm{H}}$ found in many theoretical and numerical studies  \citep[e.g.][]{martin11a,szulagyi14a}.

Moreover, we can estimate the flux of the circumplanetary disc at 1.3 mm by adopting the same approach as  \citet{zhu17a}. We consider a planet of mass $0.7 \, \mathrm{M_{\mathrm{J}}}$ at $61.7$ au from the central star accreting at a rate of $2.5\, \times 10^{-5} \,\mathrm{\mathrm{M_{\mathrm{J}}}\, yr^{-1}}$, surrounded by a disc with an outer radius of $\sim R_{\mathrm{H}}/2=1.88$ au. 
In addition to the viscous heating, we also consider the emission given by the release of accretion energy at the disc boundary layer $L_{\mathrm{bl}}=\mathcal{G} M_{\mathrm{p}}\dot{M}_{\mathrm{p}}/2r_{\mathrm{in,cpd}}$ where $r_{\mathrm{in,cpd}}$ is the inner radius, assumed equal to the Jupiter radius \citep{armitage10a}. Moreover, we take into account both the heating from the ISM external radiation and the irradiation from the central protostar, which is the only heating term included in our radiative transfer calculations.
Since the circumplanetary disc is mostly optically thick at 1.3 mm (see Fig.~\ref{fig:tausurf}) the effective brightness temperature can be roughly estimated as follows
\begin{equation}
T_{\mathrm{b}}^4=\frac{3}{8}\frac{\kappa_{\mathrm{R}}}{\kappa_{\mathrm{mm}}}T_{\mathrm{eff}}^4+T_{\mathrm{bl}}^4+T_{\mathrm{ISM}}^4+T_{\mathrm{ext}}^4,
\label{eq:brillancetemp}
\end{equation}
where $\kappa_{\mathrm{R}}=10 \,\mathrm{cm^2\, g^{-1}}$ is the Rosseland mean opacity and $\kappa_{\mathrm{mm}}= 0.25\,\mathrm{cm^2\, g^{-1}}$ is the opacity at 1.3 mm (expressed in $\mathrm{cm^2}$ per gram of gas mass, assuming a dust-to-gas mass ratio of 0.1). 
The temperature $T_{\mathrm{eff}}$ is related to the heating produced by the dissipation of the gravitational potential energy due to the viscous accretion (Eq.~1 of \citealt{zhu17a}) while $T_{\mathrm{bl}}$ is the contribution from the boundary layer (Eq.~3 of \citealt{zhu17a}).
The external temperature $T_{\mathrm{ISM}}$ of the ISM is assumed to be equal to 10 K and the contribution in temperature $T_{\mathrm{ext}}$ given by the irradiation from the central source has been computed by the radiative transfer simulations ($\sim$ 20 K).
Starting from the effective brightness temperature expressed in Eq.~\ref{eq:brillancetemp}, the total flux at 1.3 mm emitted by the circumplanetary disc inclined by an angle of $28.5^{\circ}$ at a distance of $139$ pc is $\sim$ 1.1 $\mathrm{ mJy}$. If we take into account only the external irradiation (following our assumptions in radiative transfer simulations) we obtain a flux of $\sim$ 336 $\mathrm{\mu Jy}$ at the same wavelength. Therefore, according to the Online ALMA sensitivity calculator, a total on source integration time of $\sim$ 20 min would be enough to detect the disc at 10$\sigma$. 

Fig.~\ref{fig:cfrmodelbnads} shows the synthetic observations of our model at ALMA Bands 10, 8, 7, 6, 4 and 3, corresponding to the wavelengths of 0.35, 0.75, 0.87, 1.3, 2.1 and 3 mm, respectively.
We adopt a total on source time of 10 hours and Cycle 4 antenna configurations (alma.cycle4.6 in Band 10, alma.cycle4.8 in Bands 8 and 7 and alma.cycle4.9 in Bands 6, 4 and 3), which offer a spatial resolution high enough to resolve the circumplanetary disc. Our synthetic observations show that the circumplanetary disc is marginally detected in all the ALMA Bands, especially from Band 8 to Band 4. Contrary to what observed by \citet{szulagyi17a}, the disc mostly emits at longer wavelengths due to its lower temperature. 
However, as already remarked, our simulations do not include any form of heating related to the disc formation and accretion onto the planet. Therefore, both the disc temperature and emission shown in our multi-wavelength simulations are underestimates of the real values at shorter ALMA wavelengths (see \citealt{szulagyi17a} for a more detailed analysis). \newtext{In other words, the brightness of the circumplanetary disc is only given by the reprocessed light coming from the circumstellar disc directly irradiated by the central source.} 






\subsection{Planet origin}
The recent detection of gaps and rings in protoplanetary discs suggests that small-scale axisymmetric dust structures are common and may be interpreted as a result of ubiquitous processes in disc evolution and planet formation. Regardless of the mechanism behind their origin, rings and gaps are found in young (HL Tau and Elias 24) and evolved (TW Hydrae, HD 163296 and HD 169142) sources. Importantly, these structures are observed at large distances ($\gtrsim$ 10 au) from their star and, although the measurement of the gap shape is mostly hampered by the resolution of the observations, their typical size is in the range 5-30 au \citep{zhang16a}. Recent studies have shown that these detections can be well reproduced by assuming the presence of planets with mass in the range $[0.1,0.8]\, \mathrm{M_{\mathrm{J}}}$ \citep{dipierro15b,isella16b,fedele17b,fedele17a}. However, assuming that these gaps are carved by planets, the question about the mechanisms behind the origin and growth of this kind of planets in young protoplanetary discs is still open.

The conventional core accretion theory fails to produce Saturn-mass planets at large distances from the star in typical disc lifetimes due to the rapid decline of planetesimal and gas density \citep[e.g.][]{helled14a,baruteau16a}.  One possible alternative to form giant planets at large orbital separation (40 - 100 au) is by the gravitational instability \citep{boss97a,clarke09a}. 
The threshold for the growth of density perturbations induced by gravitational instabilities and the exact conditions that lead to the disc breaking and formation of bound clumps is still under debate \citep{meru12a}. However, even in the most favourable circumstances, the typical masses of these clumps are of the order of $\sim10\, \mathrm{M_{\mathrm{J}}}$ \citep{kratter16a}.

One of the possible explanations for the presence of Sub-Jovian mass planets at large orbital distances is to assume that fast planetary formation occurs in the inner disc region and then the forming protoplanets undergo migrations towards outer disc regions. Recent studies have shown that outward migration can be induced by a large range of mechanisms, mostly related to the interaction between the planet and the coorbital mass (see Sect.~3.1.4 by \citealt{baruteau16a}).
In any case, if these structures are low density dust regions carved by planets, the main picture that is emerging  is that planet formation occurs much faster than expected.

\subsection{Limitations}
\label{sec:limits}
The greatest challenge in the interpretation of disc observations is to develop a single model capable of reproducing high resolution, multi-tracer and multi-wavelength observations of a given protoplanetary disc \citep{haworth16a}. 
In this paper, we focus on a part of the full disc modelling, based on computing the dynamics of gas and a pressureless dust fluid made of non-growing spherical grains in non-magnetized discs and a post-processing evaluation of the temperature structure and disc emission due to the irradiation from the central star. Although we use state of the art dynamical and radiative transfer modelling of protoplanetary discs, this approach suffers from some uncertainties that might affect our results.

 The primary source of uncertainty comes from the assumption that the disc substructures observed at \mbox{(sub-)}millimetre wavelengths are related to variations of the dust density and temperature structure across the disc, assuming a uniform choice for the shape, porosity and chemical composition of dust across all the discs. Any structure detected in optically thin disc emission can be also related to changes in the optical properties of dust grains.  Although this approach is mostly adopted in literature, a self consistent model of the structural and chemical composition of the dust is required to better compute the observational predictions of our disc models. 
 A second limitation of the model is based on the computation of the evolution of a single species of dust grains. Performing multiple grain species simulations would allow us to fully capture the aerodynamical coupling between the dust and gas and self-consistently compute the spatial distribution of solids spanning an entire grain size distribution \citep{ricci18a}. 
Thirdly, we assume that the presence of the planet does not affect the thermal structure of the gas. The temperature of the gas in the code is prescribed and not self-consistently computed.
This simple prescription for the disc thermal response, recently investigated by \citet{isella16a}, could lead to erroneous computation of the disc thermal emission around the planet location, especially for those planets able to carve a gap in the gas disc \citep{jang-condell03a,jang-condell12a}. By analyzing the results presented by \citet{isella16a}, due to the shallowness of the gap carved by the planet in our simulation, the disc is expected to be cooler in the gap by about $\sim$ 10\% and warmer at the gap outer edge by a similar factor (see Fig.~6 and 8 by \citealt{isella16a}). Such a small effect will marginally affect our results and might even reproduce better the observations (see Sect.~\ref{sect:simulaalma}). In detail, Fig.~\ref{fig:cfrmodel} and \ref{fig:cfrphotom} show that the emission coming from the gap outer edge is fainter than the real disc emission at the same location. An increase of the temperature by  $\sim$ 10\% would produce a better match between our simulation and the real image.
Moreover, we show that the mass of the planet inferred by our analysis is well below the limit planet mass needed to carve a deep gap in the gas disc. Therefore, the heating produced by the very weak shocks induced by such a low mass planet is expected to be negligible \citep{richert15a,hord17a}.
Fourthly, the contributions to the dust temperature of the heating due to gas and dust accretion onto the sinks, viscous heating and the cooling mechanisms are not included in our work. This simplistic approach might affect the observational predictions of the circumstellar and circumplanetary discs \citep{szulagyi17a,zhu17a}.
Finally, there remain uncertainties in the exact gas properties in Elias 24. Our observations do not give any direct indication on the 3D gas structure and the gas viscosity. As already remarked, the gas density (for low dust-to-gas ratio) regulates the dynamics of dust grains in protoplanetary discs and affects the planet migration and accretion. Moreover, the viscous-like gas disc evolution adopted in our simulations is still under debate in this field \citep{rafikov17a,tazzari17a} and the efficiency of the different sources of angular momentum transport in the presence of a planet is still an open question (see Sect.~6.2 of \citealt{zhu13a}). 



\section{Conclusion}
\label{sec:conclusion}

We present ALMA Cycle 2 observations of the dust thermal continuum emission from the Elias 24 protoplanetary disc at a wavelength of 1.3 mm with an angular resolution of $\sim 0.2''$ ($\sim 28$ au at $139$ pc). 
The dust continuum emission map reveals a dark and bright ring at a radial distance of $ 0.47''$ and  $0.58''$ from the central star ($\sim 65$ and $\sim 81$ au, respectively).  In the outer disc, the radial intensity profile shows a change of concavity at $ 0.71''$ and $ 0.87''$ ($\sim 99$ and $\sim 121$ au, respectively).
We explore the hypothesis  that the dust continuum emission map is shaped by the presence of a planet embedded in the disc.
We have carried out 3D SPH simulations of disc gas+dust evolution in a variety of disc models hosting an embedded migrating and accreting planet. The resulting dust density distribution is then used to carry out 3D radiative transfer calculations in order to produce synthetic ALMA observations of our disc model. By comparing our synthetic emission maps with the observations, we find that the observed features can be reasonably well reproduced by a planet with mass of $\sim 0.7 \, \mathrm{M_{\mathrm{J}}}$ at $\sim$ 62 au from the central star. 
The pile-up of dust grains in the gap outer edge and the radial gradient in the inward dust drift from the outer disc produce a double-hump feature in the dust density distribution, which is detected in the dust continuum. 

The picture that is emerging from the recent high resolution and high sensitivity observations of protoplanetary discs \citep{alma-partnership15a,andrews16a,isella16b,loomis17a,fedele17b,fedele17a} is that gap and ring-like features are prevalent in a large range of discs with different masses and ages. 
New high resolution and high fidelity ALMA images of dust thermal and CO line emission and high quality scattering data will be helpful to find further evidences of the mechanisms behind their formation.

\section*{Acknowledgements}
\newtext{We wish to thank Sebastian Stammler for fruitful discussions and acknowledge the referee for constructive comments that improved this manuscript.}
This paper makes use of the following ALMA data: ADS/JAO.ALMA\#2013.1.00498.S. ALMA is a partnership of ESO (representing its member states), NSF (USA) and NINS (Japan), together with NRC (Canada) and NSC and ASIAA (Taiwan) and KASI (Republic of Korea), in cooperation with the Republic of Chile. The Joint ALMA Observatory is operated by ESO, AUI/NRAO and NAOJ.
This research used the ALICE High Performance Computing Facility at the University of Leicester. Some resources on ALICE form part of the DiRAC Facility jointly funded by STFC and the Large Facilities Capital Fund of BIS. The authors are grateful to C.~P. Dullemond for making RADMC-3D available.
This project has received funding from the European Research Council (ERC) under the European Union's Horizon 2020 research and innovation programme (grant agreement No 681601). J.M.C. acknowledges support from the National Aeronautics and Space Administration under Grant No. 15XRP15\_20140 issued through the Exoplanets Research Program. M.T. has been supported by the DISCSIM project, grant agreement 341137 funded by the European Research Council under ERC- 2013-ADG. We used SPLASH \citep{price07a}.

\label{lastpage}

\bibliography{biblio}

\end{document}